\documentclass[letterpaper, 10 pt, conference]{ieeeconf}  % Comment this line out if you need a4paper

\IEEEoverridecommandlockouts                              % This command is only needed if 
\usepackage{graphics} % for pdf, bitmapped graphics files
\usepackage{epsfig} % for postscript graphics files
\usepackage{times} % assumes new font selection scheme installed
\usepackage{amsmath} % assumes amsmath package installed
\usepackage{amssymb}  % assumes amsmath package installed
\usepackage{paralist}
\usepackage{algorithmic,algorithm}
\usepackage{subfig}
\usepackage{xcolor}
\usepackage{comment}
\newtheorem{theorem}{Theorem}
\newtheorem{definition}{Definition}
\newtheorem{lemma}{Lemma}

\newtheorem{assumption}{Assumption}

\DeclareMathOperator*{\argmin}{arg\,min}
\title{
Safe and Quasi-Optimal Autonomous Navigation in Sphere Worlds
}

\author{Ishak Cheniouni, Abdelhamid Tayebi and Soulaimane Berkane % stops a space
	\thanks{This work was supported by the National Sciences and Engineering Research Council of Canada (NSERC), under the grants NSERC-DG RGPIN 2020-06270 and NSERC-DG RGPIN-2020-04759.} 
	\thanks{ The authors are with the Department of Electrical Engineering,  Lakehead University, Thunder Bay, ON P7B 5E1, Canada (e-mail:  cheniounii, atayebi@lakeheadu.ca)}
	\thanks{ Abdelhamid Tayebi is also with the Department of Electrical and Computer Engineering, Western University, London, ON N6A 3K7, Canada}%
	\thanks{S. Berkane is also with the Department Computer Science and Engineering,  University of Quebec in Outaouais, QC J8X 3X7, Canada (e-mail: soulaimane.berkane@uqo.ca).}%
}

\begin{document}

\maketitle
\thispagestyle{empty}
\pagestyle{empty}

%%%%%%%%%%%%%%%%%%%%%%%%%%%%%%%%%%%%%%%%%%%%%%%%%%%%%%%%%%%%%%%%%%%%%%%%%%%%%%%%
\begin{abstract}

We propose a continuous feedback control strategy that steers a point-mass vehicle safely to a desired destination, in a quasi-optimal manner, from almost all initial conditions in an $n$-dimensional Euclidean space cluttered with spherical obstacles. The main idea consists in avoiding each obstacle via the shortest path within the cone enclosing the obstacle, and moving straight towards the target when the vehicle has a clear line of sight to the target location. The proposed control strategy ensures safe navigation with almost global asymptotic stability of the equilibrium point at the target location. 2D and 3D simulation results, illustrating the effectiveness of the proposed approach, are presented.

%This avoidance approach satisfies Nagumo's condition for safety and we call it optimal avoidance maneuver. The avoidance maneuver is successively performed on the encountered obstacles till the vehicle has a line of sight to the destination where it takes a line segment to it. The vehicle reaches the destination safely from all initial positions in the free space, except for a set of measure zero. Simulation results, illustrating the effectiveness of the proposed approach, will be presented.

\end{abstract}

%%%%%%%%%%%%%%%%%%%%%%%%%%%%%%%%%%%%%%%%%%%%%%%%%%%%%%%%%%%%%%%%%%%%%%%%%%%%%%%%
\section{INTRODUCTION}
Safe autonomous navigation consists in steering a robot from an initial position to a final destination while avoiding obstacles. The existing solutions for this problem can be classified into two main approaches. The first approach is the plan-and-track approach, which consists in generating, from a map of the environment, a collision-free path to be tracked via a feedback controller. Among the path-finding algorithms, we can cite the Dijkstra algorithm \cite{Dijkstra1959ANO} or the A$^{\star}$ (A star) algorithm \cite{Astar}, which rely on grids or graphs representing the environment where the shortest path is determined. We can also find reactive motion planning algorithms such as the family of Bug algorithms \cite{Bug1,Bug2} which are used to navigate in planar environments without convergence and shortest-path generation guarantees.
%guarantees on the optimality of the generated paths. 
%However, the existing path planning based methods are somehow computationally expensive for on-board implementation.

The second approach, referred to as feedback-based approach, is a direct approach which consists in designing, in one shot, a feedback control strategy that steers the robot to the target location along a collision-free path. The direct approach can be further refined into two sub-classes: the sensor-based (reactive) class wherein the robot does not need to have an \textit{a priori} knowledge of its environment, and the class of control strategies that rely on global (or partial) \textit{a priori} knowledge of the environment. 
% The feedback-based approaches do not rely on offline searches and do not generally generate the shortest paths. 
%The existing solutions of the second approach do not generally generate optimal paths perform offline searches but instead use the workspace's available real-time information. 
The artificial potential field methods are an example. They consider a robot moving in a force field where the destination generates an attractive force, and the obstacles generate repulsive forces \cite{khatib}. The destination is the minimum of the potential function, and the negative gradient leads safely to it. These methods suffer from two problems, namely, the generation of local minima where the robot may get trapped instead of reaching the goal, and if the goal is reached the generated path is not generally the shortest collision-free path. To address the problem of local minima, the authors in \cite{k_R_90} proposed a navigation function (NF) whose negative gradient is the control law that steers the robot from almost all initial conditions to the target location in an \textit{a propri} known sphere world.
%which allows to design  potential field with a unique minimum at the destination position and the remaining critical points as saddles. 
In order to navigate in more general spaces, diffeomorphisms from sphere worlds to more complex worlds were proposed in \cite{Pr_K_R_91,R_k_92}. 
%Besides the navigation function, the diffeomorphisms must also be tuned offline. However, even if the existence of the parameters is theoretically guaranteed, their calculation is not simple \cite{LoizouNT4}. 
The authors in \cite{LoizouNT1, LoizouNT3} proposed tuning-free navigation functions and  diffeomorphisms from a point world to sphere world or a star world. A sufficient condition was given in \cite{SnsNF6} for an artificial potential to be a navigation function in environments containing smooth, non-intersecting, and strongly convex obstacles. More recently, a tuning-free navigation function based on harmonic functions has been proposed in \cite{loizou2021correctbyconstruction} for sensor-based autonomous navigation.

In \cite{Arslan2019}, the authors proposed a new sensor-based autonomous navigation strategy (different from the NF-based approach) by constructing a compact obstacle-free local set around the robot using the hyperplanes separating the robot from the neighboring obstacles and then steering the robot towards the projection of the target location onto the boundary of this compact set. This approach ensures safe navigation through unknown strongly convex obstacles and convergence to the destination from everywhere, except from a set of zero Lebesgue measure. This work has been extended for non-convex star-shaped obstacles in \cite{vasilopoulos1}, and polygonal obstacles with possible overlap in \cite{Vasilopoulos2}.
%, and in \cite{Vasilopoulos3}, the target is considered mobile.

A sensor-based autonomous navigation approach, relying on Nagumo's theorem  \cite{Nagumo} and using tangent cones, was proposed in \cite{souVeloCones}. This approach guarantees safety through an appropriate switching between a stabilizing controller and an obstacle avoidance controller. Control Barrier Functions (CBFs) and Control Lyapunov Functions (CLFs) were used in \cite{Barrier6, Barrier7} along with a quadratic program to design navigation controllers ensuring the stabilization of the desired target location with safety guarantees.

%In \cite{Barrier2,Barrier3}, Nagumo's theorem was used along with barrier functions to verify the safety of nonlinear and hybrid systems. By analogy to Control Lyapunov Functions (CLFs), Control Barrier Functions (CBFs) have been introduced in \cite{Barrier4} to design feedback controllers with safety guarantees for dynamical systems. CBFs and CLFs were unified in \cite{Barrier6, Barrier7} through quadratic programs to design navigation controllers guaranteeing the stabilization of the desired target location with safety guarantees.

None of the aforementioned work has achieved global asymptotic stability of the target location due to the topological obstruction pointed out in \cite{k_R_90}. To overcome this problem, a hybrid state feedback control strategy, with robust global asymptotic stabilization of a target location, was proposed in \cite{hybr1} in the case of a single obstacle. %Following the same approach, a hybrid multi-agent autonomous navigation method for a group of vehicles in an environment containing a single obstacle was proposed in \cite{hybr2}. 
%A hybrid state feedback, guaranteeing robust global asymptotic stability of a target location while avoiding the neighborhood of a point obstacle, was proposed in \cite{hybr3}, and extended in \cite{hybr4} to the case of multiple obstacles. 
A hybrid feedback controller, with global asymptotic stability guarantees, has been proposed in \cite{SoulaimaneHybTr} for a vehicle navigating in an $n$-dimensional Euclidean space filled with ellipsoidal obstacles, and in \cite{Mayur2022} for robots navigating in two-dimensional environments filled with arbitrary convex obstacles.  

While safe global (or almost global) convergence to a target is achieved in environments with specific geometries, all the feedback-based approaches mentioned above do not generally generate the shortest collision-free paths. In this paper, we address this problem by proposing a continuous quasi-optimal\footnote{This term will be rigorously defined later.} feedback control strategy guaranteeing safe navigation from almost all initial conditions, in a sphere world, to the target location while generating quasi-optimal collision-free paths. Our approach relies on an iterative projection strategy that generates a feedback control law leading to successive locally optimal collision-free paths with respect to the successive obstacles in the robots path.

%maneuver to avoid an obstacle by taking the shortest path within its enclosing cone and moving straight towards the target when the vehicle has a clear line of sight to the target location. This maneuver is used successively to avoid the visited obstacles, which generates a quasi-optimal trajectory and ensures safe navigation with almost global asymptotic stability of the equilibrium point at the target location.

\section{Notations and Preliminaries}
Throughout the paper, $\mathbb{N}$, $\mathbb{R}$ and $\mathbb{R}_{>0}$ denote the set of natural numbers, real numbers and positive real numbers, respectively. The Euclidean space and the unit $n$-sphere are denoted by $\mathbb{R}^n$ and $\mathbb{S}^n$, respectively. The Euclidean norm of $x\in\mathbb{R}^n$ is defined as $\|x\|:=\sqrt{x^\top x}$ and the angle between two non-zero vectors $x,y\in\mathbb{R}^n$ is given by $\angle (x,y):=\cos^{-1}(x^\top y/\|x\|\|y\|)$ . The Jacobian matrix of a vector field $f: \mathbb{R}^n\rightarrow\mathbb{R}^n$ is given by $J_x(f(x))=[\nabla_x f_1\dots\nabla_x f_n]^\top$ where $\nabla_x f_i=[\frac{\partial f_i}{\partial  x_1}\,\dots\,\frac{\partial f_i}{\partial  x_n}]^\top$ is the gradient of the $i$-th element $f_i$. %The successive composition, with respect to the first variable, of the map $g: \mathbb{R}^n\times\mathbb{R}^n\times\mathbb{N}\setminus{0}\rightarrow\mathbb{R}^n$ is defined as $$\mathop{\bigcirc}_{p=1}^{\bar{p}}g(x,y,p)=\underbrace{g\circ\dots\circ g(x,y,1)}_{\bar{p}\;\textrm{times}},$$ where $\bar{p}\in\mathbb{N}\setminus{0}$, and $g\circ g(x,y,p)=g(g(x,y,p),y,p+1)$.
We define a ball centered at $x\in\mathbb{R}^n$ and of radius $r\in\mathbb{R}_{>0}$ by the set $\mathcal{B}(x,r):=\left\{q\in\mathbb{R}^n|\;\|q-x\| \leq r\right\}$. The interior and the boundary of a set $\mathcal{A}\subset\mathbb{R}^n$ are denoted by $\mathring{\mathcal{A}}$ and $\partial\mathcal{A}$, respectively. The relative complement of a set $\mathcal{B}\subset\mathbb{R}^n$ with respect to a set $\mathcal{A}$ is denoted by $\mathcal{B}^c_\mathcal{A}$. The distance of a point $x\in\mathbb{R}^n$ to a closed set $\mathcal{A}$ is defined as $d(x,\mathcal{A}):=\min\limits_{q\in\mathcal{A}}\|q-x\|$. The cardinality of a set $\mathcal{N}\subset\mathbb{N}$ is denoted by $\mathbf{card}(\mathcal{N})$. The line segment connecting two points $x,y\in\mathbb{R}^n$ is defined as $\mathcal{L}_s(x, y):=\left\{q\in\mathbb{R}^n|q=x+\delta(y-x),\;\delta\in[0,1]\right\}$. The parallel and orthogonal projections are defined as follows:
\begin{align}
    \pi^{\parallel}(v):=vv^\top,\quad\pi^{\bot}(v)&:=I_n-vv^\top,
\end{align}
where $I_n\in\mathbb{R}^{n\times n}$ is the identity matrix and $v\in\mathbb{S}^{n-1}\setminus\{0\}$. Therefore, for any vector $x$, the vectors $\pi^{\parallel}(v) x$ and $\pi^{\bot}(v) x$ correspond, respectively, to the projection of $x$ onto the line generated by $v$ and onto the hyperplane orthogonal to $v$.
\begin{comment}
For $v\in\mathbb{S}^{n-1}\setminus\{0\}$, one has the following useful properties \cite{SoulaimaneHybTr}
\begin{align}
\pi^{\parallel}(v)\pi^{\parallel}(v)&=\pi^{\parallel}(v),           &  \pi^{\parallel}(v)v &=v,\\
\pi^{\bot}(v)\pi^{\bot}(v)&=\pi^{\bot}(v),         &  \pi^{\bot}(v)v&=0.
\end{align}
\end{comment}
A conic subset of $\mathcal{A}\subseteq\mathbb{R}^n$, with vertex $x\in\mathbb{R}^n$, axis $a\in\mathbb{R}^n$, and aperture $2\psi$ is defined as follows \cite{HybBerkaneECC2019}:
\begin{align}
    \mathcal{C}^{\Delta}_{\mathcal{A}}(x,a,\psi):=\left\{q\in\mathcal{A}|\|a\|\|q-x\|\cos(\psi)\Delta a^\top(q-x)\right\},
\end{align}
where $\psi\in(0,\frac{\pi}{2}]$ and $\Delta\in\left\{\leq,<,=,>,\geq\right\}$,  with $``="$, representing the surface of the cone, $``\leq"$ (resp. $``<"$) representing the interior of the cone including its boundary (resp. excluding its boundary), and $``\geq"$ (resp. $``>"$) representing the exterior of the cone including its boundary (resp. excluding its boundary). The set of vectors parallel to the cone $\mathcal{C}^=_{\mathbb{R}^n}(x,a,\psi)$ is defined as follows:
    \begin{align}\label{17}
        \mathcal{V}(a,\psi):=\left\{v\in\mathbb{R}^n|\;\;v^\top a=\|v\|\|a\|\cos(\psi)\right\}.
    \end{align}
%%%%%%%%%%%%%%%%%%%%%%%%%%%%%%%%%%%%%%%%%%%%%%%%%%%%%%%%%%%%%%%%%%%%%%%%%%%%%%%%%%%%%%%%%%%%%%%%
\section{Problem Formulation}
We consider 
a point mass vehicle  $x\in\mathbb{R}^n$ moving inside a spherical workspace $\mathcal{W}\subset\mathbb{R}^n$ centered at the origin $0$ and punctured by $m\in\mathbb{N}$ open balls $\mathcal{O}_i$ such that
\begin{align}
\mathcal{W}:=\mathcal{B}(0,r_0),\quad   \mathcal{O}_i:=\mathcal{B}(c_i,r_i),\;\;i\in\mathbb{I}:=\{1,\dots,m\},\label{m2}
\end{align}
where $r_0>r_i>0$ for all $i\in\mathbb{I}$. The free space is, therefore, given by the closed set
\begin{align}\label{8}
   \mathcal{F}:=\mathcal{W}\setminus\bigcup\limits_{i=1}^{m}\mathring{\mathcal{O}}_i.
\end{align}
For $\mathcal{F}$ to be a valid sphere world, as defined in \cite{k_R_90}, the obstacles $\mathcal{O}_i$ must satisfy the following assumptions:
\begin{assumption}\label{as:1}
 The obstacles are completely contained within the workspace and separated from its boundary, {\it i.e.,}
 \begin{align}\label{9}
    \|c_i\|+r_i<r_0,\;\;\forall i\in\mathbb{I}.
\end{align}
\end{assumption}
\begin{assumption}\label{as:2}
 The obstacles are disjoint, {\it i.e.,}
 \begin{align}\label{10}
    \|c_i-c_j\|>r_i+r_j,\;\;\forall i,j\in\mathbb{I},\,i\neq j.
\end{align}
\end{assumption}
Consequently, the boundary of the free space $\mathcal{F}$ is given as follows:
\begin{align}\label{11}
    \partial\mathcal{F}:=\partial\mathcal{W}\bigcup\Bigl(\bigcup\limits_{i=1}^{m}\partial\mathcal{O}_i\Bigr).
\end{align}
We consider the following first-order vehicle dynamics
\begin{align}\label{12}
    \Dot{x}=u,
\end{align}
where $u$ is the control input. The objective is to determine a continuous Lipschitz state-feedback controller $u(x)$ that safely steers the vehicle from almost any initial position $x(0)\in\mathcal{F}$ to any given desired destination $x_d\in\mathcal{F}$. In particular, the closed-loop system
\begin{equation}\label{eq:closed-loop-system}
    \dot x=u(x),\quad x(0)\in\mathcal{F}
\end{equation}
must ensure forward invariance of the set $\mathcal{F}$, almost global asymptotic stability of the equilibrium $x=x_d$, and  {\it quasi-optimal obstacle avoidance maneuver}. A quasi-optimal obstacle avoidance maneuver is defined as follows.
\begin{comment}
must ensure forward invariance of the set $\mathcal{F}$, almost global asymptotic stability of the equilibrium $x=x_d$, and local {\it optimal obstacle avoidance maneuvers}. An optimal local obstacle avoidance maneuver ({\it i.e.,} with respect to a given single obstacle) is defined as follows:

\begin{definition}\label{def1}
A trajectory $x(t)$ is said to generate an optimal obstacle avoidance maneuver between $x_0$ and $x_d$, with respect to a given obstacle $\mathcal{O}$, if it coincides with the euclidean shortest path, {\it i.e.,} the integral\footnote{The notation $|x^\prime|(t)$ denotes the metric derivative of $x(t)$ at $t$. For continuously differentiable paths, $|x^\prime|(t)$ coincides with $||\dot x(t)||$. }
{\color{blue}\begin{align}
    \int_0^{T}|x^\prime|(t)dt
\end{align}
is minimum across all possible safe trajectories $x(t)$ where $T \geq0$}. 

%, for each time $t>0$, the direction of the velocity vector $\dot{x}(t)$ must not cross the obstacle. At the same time, the vehicle remains as close as possible to its destination.
Let $x_i(t)$, $i\in \mathcal{N}\subseteq\mathbb{N}$, be a set of feasible (collision-free) trajectories such that $x_i(0)=x_0$ and $x_i(t_f^i)=x_d$. The trajectory $x_m(t)$ is said to be optimal (shortest Euclidean path from $x_0$ to $x_d$) if $\argmin_{i\in \mathcal{N}} \int_0^{t_f^i}|x_i^\prime|(t)dt=m$. Note that the notation $|x_i^\prime|(t)$ denotes the metric derivative of $x_i(t)$ at $t$. For continuously differentiable paths, $|x_i^\prime|(t)$ coincides with $||\dot x_i(t)||$. 
\end{definition}
\end{comment}

Let $x(t)$ be the generated trajectory of the closed-loop system \eqref{eq:closed-loop-system}, such that $\lim_{t \to \infty} x(t) = x_d$. For some $\varepsilon\geq0$ and for each $i\in\mathcal{VO}\subseteq \mathbb{I}$, where $\mathcal{VO}$ is the ordered list of visited obstacles, let $x_{d,\varepsilon}^i$ denote the point where the curve $x(t)$ leaves the ball $\mathcal{B}(c_i,r_i+\varepsilon)$ enclosing the  $i$th obstacle, with $x_{d,\varepsilon}^z:=x_d$ and $z=\mathbf{card}(\mathcal{VO})$. Also, let $x_{0,\varepsilon}^i:=x_d^{i-1}$ with $x_{0,\varepsilon}^1:=x(0)$.

\begin{definition}\label{def2}
The trajectory $x(t)$ is said to be generated by a quasi-optimal obstacle avoidance maneuver if there exists $\varepsilon\geq 0$ such that the local obstacle avoidance maneuvers, with respect to each obstacle $i$ and between $x_{0,\varepsilon}^i$ and $x_{d,\varepsilon}^i$, are all optimal (\textit{i.e.,} they generate the shortest collision-free Euclidean paths). 
\end{definition}
%%%%%%%%%%%%%%%%%%%%%%%%%%%%%%%%%%%%%%%%%%%%%%%%%%%%%%%%%%%%%%%%%%%%%%%%%%%%%%%%%%%%%%%%%%%%%%
\section{Sets Definition and Obstacles Classification}\label{section:sets}
\begin{figure}[h!]
\centering
\includegraphics[scale=0.3]{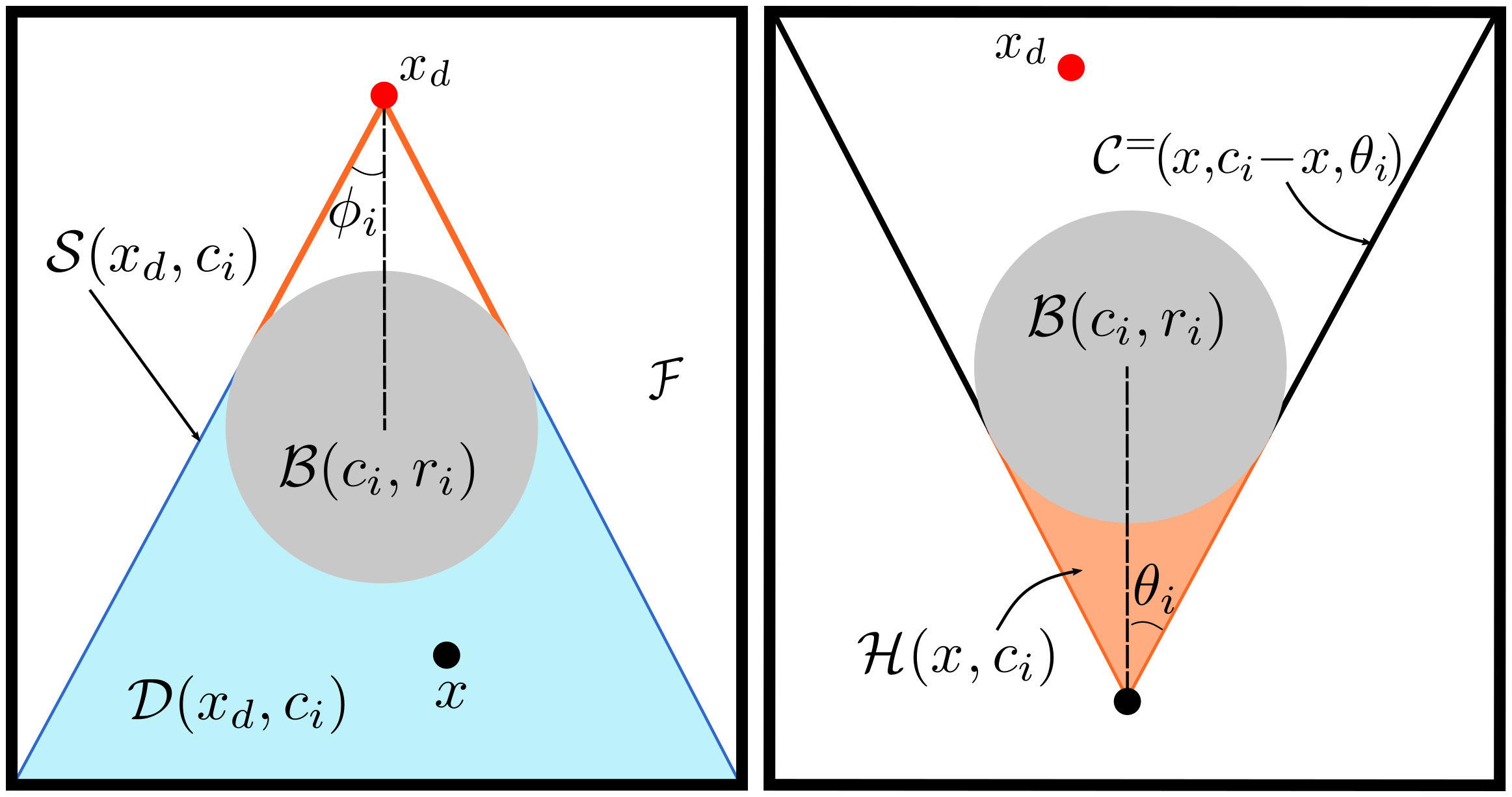}
\caption{2-D representation of the sets in Section \ref{section:sets}. 
% The left figure is the interior of the cone with vertex $x_d$, axis $(c_i-x_d)$ and aperture $2\phi_i$. The blue region represents the shadow region, the thick blue lines represent the surface separating the shadow region and its complement in the free space, and the orange region represents the hat of the cone. The left figure is the cone
% with vertex $x$, axis $(c_i-x)$ and aperture $2\theta_i$. The Orange region represents the hat of the cone.
}
\label{fig:fig1}
\end{figure}

\begin{figure}[h!]
\centering
\includegraphics[scale=0.25]{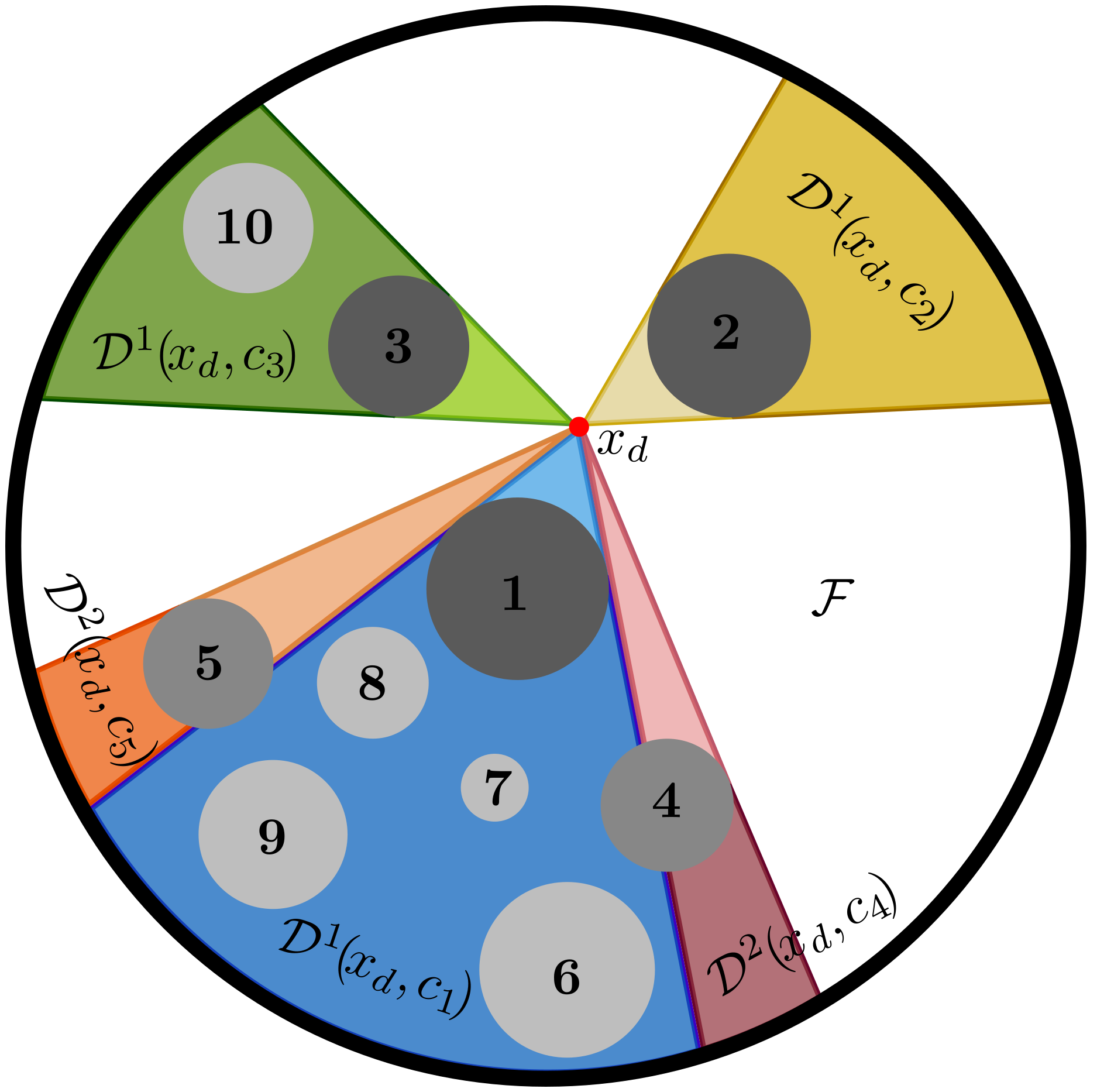}
\caption{Sub-shadow regions of obstacles from different classes in a 2-D sphere world. First-generation obstacles are in dark gray, second-generation obstacles are in medium gray, and zero-generation obstacles are in light gray.}
\label{fig:fig5}
\end{figure}

In this section, we define the subsets of the free space that are needed for our proposed control design in Section \ref{section:control-design}. These are given as follows:
\begin{itemize}
\begin{comment}
\item The hat of the cone $ \mathcal{C}^{\leq}_{\mathcal{F}}(x_d,c_i-x_d,\phi_i)$ is defined as follows (orange region in Fig. \ref{fig:fig1} (left)):
     \begin{multline}\label{17}
        \mathcal{H}(x_d,c_i):=\Bigl\{q\in\mathcal{C}^{\leq}_{\mathcal{F}}(x_d,c_i-x_d,\phi_i)|\\(c_i-q)^{\top}(x_d-q)<0\Bigr\},
    \end{multline}
    where the angle $\phi_i \in(0,\frac{\pi}{2}]$ is given by
$
    \phi_i(x):=\arcsin\left(r_i/\|c_i-x_d\|\right).
$
\end{comment}
    \item The shadow region: the area where the vehicle does not have a clear line of sight to the target is defined as follows (blue region in Fig. \ref{fig:fig1} (left)):
    \begin{multline}\label{18}
        \mathcal{D}(x_d,c_i):=\Bigl\{q\in\mathcal{C}^{\leq}_{\mathcal{F}}(x_d,c_i-x_d,\phi_i)|\\(c_i-q)^{\top}(x_d-q)\geq0\Bigr\},
    \end{multline}
    where the angle $\phi_i \in(0,\frac{\pi}{2}]$ is given by
$
    \phi_i:=\arcsin\left(r_i/\|c_i-x_d\|\right).
$
    \item The exit set separates the set $\mathcal{D}(x_d,c_i)$ and its complement with respect to $\mathcal{F}$ and is defined as follows (thick blue lines in Fig. \ref{fig:fig1} (left)):
    \begin{multline}\label{20}
        \mathcal{S}(x_d,c_i):=\Bigl\{q\in\mathcal{C}^=_{\mathcal{F}}(x_d,c_i-x_d,\phi_i)|\\(c_i-q)^\top(x_d-q)\geq0\Bigr\}.
    \end{multline}
    \item The hat of the cone $\mathcal{C}^{\leq}_{\mathcal{F}}(x,c_i-x,\theta_i(x))$ is defined as follows (see Fig. \ref{fig:fig1} (right)):
    \begin{multline}\label{19}
    \mathcal{H}(x,c_i):=\Bigl\{q\in\mathcal{C}^{\leq}_{\mathcal{F}}(x,c_i-x,\theta_i(x))|\;(c_i-q)^\top(x-q)\\\leq0\Bigr\},
    \end{multline}
    where the angle $\theta_i(x) \in(0,\frac{\pi}{2}]$ is given by
$
    \theta_i(x):=\arcsin\left(r_i/\|c_i-x\|\right).
$
\end{itemize}
Next, we classify the obstacles according to their visibility from the destination into different generations. An obstacle that can be fully seen from the destination is a first-generation obstacle (dark gray obstacles $\{1,2,3\}$ in Fig. \ref{fig:fig5}). A second-generation obstacle can be partially seen from the destination and is partially included in the shadow regions of the first-generation obstacles (medium gray obstacles $\{4,5\}$ in Fig. \ref{fig:fig5}). An obstacle is said to be of generation $j\geq 2$ if it is partially seen from the destination and is included in the shadow region of at least one obstacle of generation $j-1$. An obstacle that is completely hidden from the destination, whose shadow region is entirely included in the shadow regions of other obstacles, is classified as a zero-generation obstacle (light gray obstacles in Fig. \ref{fig:fig5}). Now, we define the sets related to the obstacle classification as follows:
\begin{itemize}
    \item The sub-shadow region of an obstacle $\mathcal{O}_i$ is defined as follows (see Fig.~\ref{fig:fig5}):
    \begin{align}
    \small
       % \mathcal{D}^1(x_d,c_i)&=\mathcal{D}^0(x_d,c_i):= \mathcal{D}(x_d,c_i),\label{m15}\\
        \mathcal{D}^j(x_d,c_i):=\mathcal{D}(x_d,c_i)\setminus\bigcup\limits_{l=1}^{j-1}\left[\bigcup\limits_{k\in J_i^{l}}\mathcal{D}^{l}(x_d,c_k)\right],\label{m16}
    \end{align}
    for $j\geq2$ where $J_i^j=\{k\in\left\{1,\dots,m\right\}| k\neq i\;\text{and}\;\mathcal{D}(x_d,c_i)\cap \mathcal{D}^j(x_d,c_k)\allowbreak\neq\varnothing\}$ is the set of the $j$-generation obstacles that include obstacle $i$ in their sub-shadow regions and $\mathcal{D}^1(x_d,c_i)=\mathcal{D}^0(x_d,c_i):= \mathcal{D}(x_d,c_i)$.
    \item The blind set is a subset of $\mathcal{F}$ where there is no line of sight to the destination, and it is defined as
    \begin{align}\label{m19}
        \mathcal{BL}:=\bigcup\limits_{j=1}^{s}\left[\bigcup\limits_{i\in\mathcal{G}_j}\mathcal{D}^j(x_d,c_i)\right],
    \end{align}
    where $\mathcal{G}_j$ is the set of obstacles of generation $j\in\{1,\dots,s\}$ and $s\leq m$ is the total number of generations in the workspace. 
    \item The visible set is defined as the complement of the blind set with respect to the free space $\mathcal{VI}:= \mathcal{BL}^c_\mathcal{F}.$
\end{itemize}
%%%%%%%%%%%%%%%%%%%%%%%%%%%%%%%%%%%%%%%%%%%%%%%%%%%%%%%%%%%%%%%%%%%%%%%%%%%%%%%%%%%%%%%%%%
\section{Control Design}\label{section:control-design}
\subsection{Single Obstacle Case}
We start by considering a single obstacle $\mathcal{O}_i$ and ignoring all others. We design a preliminary control law for the single obstacle case, which will be used as a baseline in the multiple obstacles case. First, in the case where the path is clear (\textit{i.e.,} $x$ belongs to the visible set  $\mathcal{VI}$), the vehicle follows a straight line to the destination under the control law $u_d(x)=-\gamma(x-x_d)$ where $\gamma \in\mathbb{R}_{>0}$. Next, in the case where the path is not clear (\textit{i.e.,} $x\in \mathcal{D}(x_d,c_i)$), we generate a control input (vehicle's velocity) that is in the direction of the cone enclosing the obstacle $\mathcal{C}^{=}_{\mathcal{F}}(x,c_i-x,\theta_i)$ while ensuring that the control input is equal to $u_d(x)$ at the exit set $\mathcal{S}(x_d,c_i)$. In particular, the direction of the control input should minimize the angle between the nominal control direction, given by $(x_d-x)$, and the set of all vectors parallel to the enclosing cone, {\it i.e.,}   the control input should belong to the set
    \begin{equation}\label{min}
    \mathcal{U}_1(x):=\mathrm{arg}\min\limits_{v_i\in\mathcal{V}(c_i-x,\theta_i)}\angle(x_d-x,v_i).
    \end{equation}
    Moreover, to ensure continuity of the control input, we impose further that the control input belongs to the set
    \begin{equation}
        \mathcal{U}_2(x):=\{v\in\mathbb{R}^n:  v=u_d(x),\; x\in\mathcal{S}(x_d,c_i)\}.
    \end{equation}
    These two conditions can be written as follows
    \begin{align}\label{L2}
        u(x)\in\mathcal{U}_1(x)\cap\mathcal{U}_2(x).
    \end{align}
    In the following lemma we show that the set $\mathcal{U}_1(x)\cap\mathcal{U}_2(x)$ is a singleton and we provide the unique solution.
    % by projecting $x_d-x$ onto the cone enclosing the obstacle $\mathcal{C}^{=}(x,c_i-x,\theta_i)$ and minimizing the angle between $x_d-x$ and $v_i$ with $v_i\in\mathcal{V}(c_i-x,\theta_i)$. This projection process can be summarized in the following minimization problem,
    % \begin{align}\label{22}
    %     u(x)=\mathrm{arg}\min\limits_{v_i\in\mathcal{V}(c_i-x,\theta_i)}\angle(x_d-x,v_i),
    % \end{align}
    %\begin{align}\label{22}
     %   \Bar{v}_i=\mathrm{arg}\min\limits_{v_i\in\mathcal{C}^=(x,c_i-x,\theta_i)}\angle(V_d,v_i),
    %\end{align}
    % with $u(x)=u_d(x)$ when $x\in\mathcal{S}(x_d,c_i)$.
\begin{lemma}\label{lem1}
Set $\mathcal{U}_1(x)\cap\mathcal{U}_2(x)$ is a singleton and the unique element is given by
\begin{align}
    u(x)=\xi(u_d(x),x,i),
\end{align}
where $\xi:\mathbb{R}^n\times\mathbb{R}^n\times\mathbb{N}\setminus{0}\to\mathbb{R}^n$ is given by
    \begin{multline}\label{alg}
   \xi(u,x,i):=\frac{\sin(\beta_i(u,x))\sin^{-1}(\theta_i(x))}{\cos(\theta_i(x)-\beta_i(u,x))}\pi^{\parallel}(\bar\xi)u,
    \end{multline}
    with
    \begin{align*}
        &\bar\xi:=\frac{\sin(\theta_i(x))u}{\sin(\beta_i(u,x))\|u\|}-\frac{\sin(\theta_i(x)-\beta_i(u,x))}{\sin(\beta_i(u,x))}\frac{(c_i-x)}{\|c_i-x\|},\\
        &\beta_i(u,x):=\angle(u,c_i-x)\leq\theta_i(x).
    \end{align*}
% \begin{align}
%   % u(x)&=\gamma\|x_d-x\|\left(V_d-\frac{\sin(\theta_i-\beta_i)}{\sin(\theta_i)}V_{ci}\right)\\
%   u(x)&=\frac{\sin(\beta_i)}{\cos(\theta_i-\beta_i)\sin(\theta_i)}\pi^{\parallel}(\bar{u}_i)u_d(x),\;\;\;\forall x\in\mathcal{D}(x_d,c_i),
% \end{align}
%  where $\bar{u}_i=\frac{\sin(\theta_i)}{\sin(\beta_i)}V_d-\frac{\sin(\theta_i-\beta_i)}{\sin(\beta_i)}V_{ci}$ is a unit vector, $V_d=\frac{x_d-x}{\|x_d-x\|}$, $V_{ci}=\frac{c_i-x}{\|c_i-x\|}$,  and $\beta_i=\angle(V_{ci},V_d)$ with $0\leq\beta_i\leq\theta_i$.
\end{lemma}
\begin{proof}
See Appendix \ref{appendix:Lemma 1}.
\end{proof}
In other words, Lemma \ref{lem1} shows that,  when $x\in\mathcal{D}(x_d,c_i)$, the control $u(x)$ is a scaled parallel projection of the nominal controller $u_d(x)$ in the direction of $\bar\xi$ which represents a unit vector on the cone enclosing the obstacle.
Finally, we obtain the following control strategy in the case of a single obstacle
\begin{align}
   u(x)= \begin{cases}\label{25}
      \displaystyle u_d(x), & x\in\mathcal{VI},\\
      \displaystyle\xi(u_d(x),x,i), & x\in\mathcal{D}(x_d,c_i).
    \end{cases} 
\end{align}%
Note that, during the avoidance maneuver, the controller depends on three arguments: the nominal control $u_d(x)$, the current position of the vehicle $x$, and the obstacle index $i$. Moreover, the trajectory of the closed-loop system \eqref{12}-\eqref{25} generates an optimal obstacle avoidance maneuver as shown in the following lemma.
\begin{lemma}\label{lem6}
The path generated by the closed-loop system \eqref{12}-\eqref{25} is the shortest path to the destination $x_d$ from every initial condition $x(0)\in\mathcal{F}\setminus\mathcal{L}_d(x_d,c_i)$ where $\mathcal{L}_d(x_d,c_i):=\left\{q\in\mathcal{D}(x_d,c_i)|\;\;q=c_i+\delta(c_i-x_d),\;\delta\in\mathbb{R}_{>0}\right\}$. 
\end{lemma}

\begin{proof}
See Appendix \ref{appendix:Lemma 6}.
\end{proof}
\begin{comment}
\begin{figure}[h!]
\centering
\includegraphics[scale=0.3]{images/fig3.png}
\caption{Representation of the closest control to the nominal control $u_d$ among the cone enclosing the obstacle $\mathcal{O}_i$ satisfying the obstacle avoidance.}
\label{fig:fig3}
\end{figure}
\end{comment}
%%%%%%%%%%%%%%%%%%%%%%%%%%%%%%%%

\subsection{Multiple Obstacles Case}
\begin{figure}[h!]
    \centering
    \includegraphics[scale=0.33]{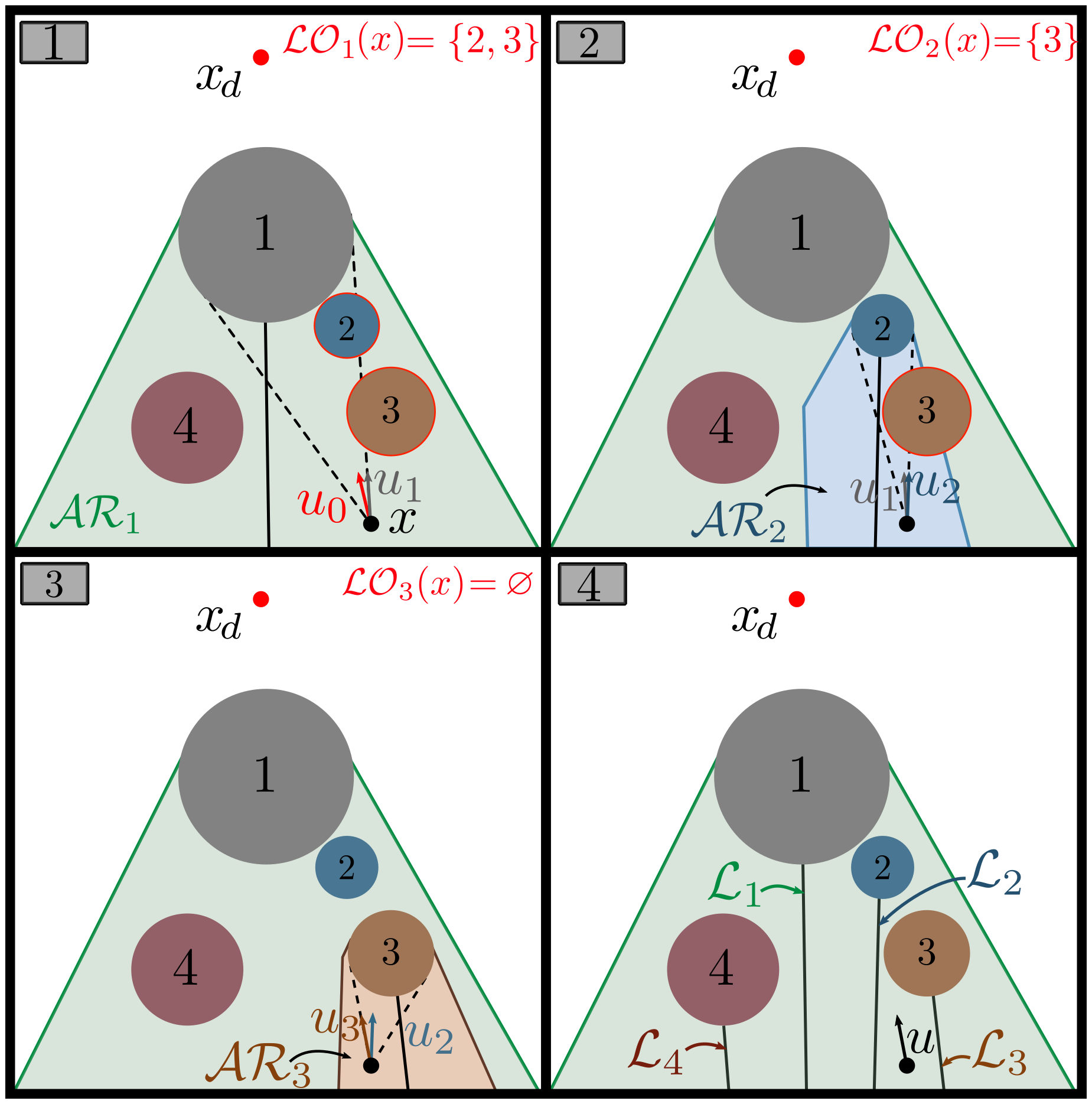}
    \caption{Successive projections of the control $u_d$ in a two-dimensional sphere world with four obstacles.}
    \label{fig:fig8}
    \end{figure}

In the case of multiple obstacles and when $x\in\mathcal{BL}$, we proceed with multiple projections as described hereafter. When $x\in\mathcal{BL}$, there exist $j\in\{1,\dots,s\}$ and $i\in\mathcal{G}_j$ such that $x\in\mathcal{D}^j(x_d,c_i)$ (definition \eqref{m19}). In this case, the obstacle $\mathcal{O}_i$ is the first to be considered, and $u_d(x)$ is projected onto its enclosing cone in a similar way as in \eqref{25}. The resulting control vector is denoted by $u_1(x)$. The next obstacle to be considered is selected from the set of obstacles shadowed by obstacle $\mathcal{O}_i$ defined as $\mathcal{LO}_i(x):=\left\{k\in\left\{1,\dots,m\right\}\setminus\left\{i\right\}|\exists q\in\partial\mathcal{O}_k,\,q\in\mathcal{H}(x,c_i) \right\}$, see Fig. \ref{fig:fig8}. Amongst obstacles in $\mathcal{LO}_i(x)$ that contain $u_1$ in their enclosing cones, we choose the closest in terms of the Euclidean distance to $\mathcal{O}_i$. If $\mathcal{LO}_i(x)=\varnothing$ or no obstacle contains $u_1$ in its enclosing cone, the path is free. Otherwise, $u_1$ will be considered as $u_d$ for the new selected obstacle and the same approach is followed to obtain $u_2$. We say that $\mathcal{O}_i$ is an \textbf{ancestor} to the selected obstacle and we repeat the selection and projection until the path is free (Fig. \ref{fig:fig8}). The obstacles selected during the successive projections at a position $x$, are grouped in an ordered list $\mathcal{I}(x)\subset\mathbb{I}$ from the first obstacle ($\mathcal{O}_i$, such that $x\in\mathcal{D}^j(x_d,c_i)$) to the last one (obstacle involved in the last projection). Let $h(x)=\mathbf{card}(\mathcal{I}(x))$ be the number of required projections at  position $x$. We define the bijection $\iota_x :\{1,\dots,h(x)\}\rightarrow\mathcal{I}(x)$ which associates to each projection $p\in\{1,\dots,h(x)\}$ the corresponding obstacle $\iota_x(p)\in\mathcal{I}(x)$. The set of positions involving obstacle $\iota_x(p)$ in the successive projections is called active region and defined as $\mathcal{AR}_k:=\left\{q\in\mathcal{BL}|k\in\mathcal{I}(q)\right\}$ with $k=\iota_x(p)$. To sum up, the intermediary control at a step $p\in\{1,\dots,h(x)\}$ and position $x\in\mathcal{AR}_{\iota_x(p)}$ is given by the recursive formula
    \begin{align}\label{eq:recursive-control}
        u_p(x)=\xi(u_{p-1}(x),x,\iota_x(p)),
    \end{align}
    with $u_0(x)=u_d(x)$ and $\xi(\cdot,\cdot,\cdot)$ as defined in Lemma \ref{lem1}.
%     where $\xi:\mathbb{R}^n\times\mathbb{R}^n\times\mathbb{N}\to\mathbb{R}^n$ is a nonlinear projection operator given by
%     \begin{multline}\label{alg}
%   \xi(u,x,k):=u-\|u\|\frac{\sin(\theta_{k}(x)-\beta_{k}(u,x))}{\sin(\theta_{k}(x))\|c_{k}-x\|}(c_{k}-x),
%      \end{multline}
%      with $\beta_k(u,x):=\angle(u,c_k-x)$.
%      Using straightforward manipulations, this projection operation at step $p$ can be also rewritten
%     \begin{align}
%     u_p(x)&=\frac{\sin(\beta_{\iota(p)})\sin^{-1}(\theta_{\iota(p)})}{\cos(\theta_{\iota(p)}-\beta_{\iota(p)})}\pi^{\parallel}(\bar{u}_{\iota(p)})u_{p-1}(x),
%     \end{align}
%     where $\bar{u}_{\iota(p)}=\frac{\sin(\theta_{\iota(p)})}{\sin(\beta_{\iota(p)})}V_p-\frac{\sin(\theta_{\iota(p)}-\beta_{\iota(p)})}{\sin(\beta_{\iota(p)})}V_{c{\iota(p)}}, V_{ci}=(c_i-x)/\|c_i-x\|$ and $V_p=u_{p-1}(x)/\|u_{p-1}(x)\|$.
Finally, the proposed control law is obtained by performing $h(x)$ successive projections and is given by
\begin{align}\label{36}
   u(x)= \begin{cases}
      u_d(x), & x\in\mathcal{VI},\\
      %\left(\prod\limits_{i\in\mathcal{I}(x)}\pi^{||}(v_i)\right)u_d(x) &
      u_{h(x)}(x),&
      x\in \mathcal{BL}.
    \end{cases} 
\end{align}
    \begin{comment}
    To sum up, the intermediary control at a step $p$ is given by
    \begin{align}\label{alg}
    %u_p(x)=\|u_{p-1}\|\Bigl(V_p-\frac{\sin(\theta_i-\beta_i)}{\sin(\theta_i)}V_{ci}\Bigr),\\
    u_p(x)=\frac{\sin(\beta_i)}{\cos(\theta_i-\beta_i)\sin(\theta_i)}\pi^{\parallel}(\bar{u}_i)u_{p-1}(x)
    \end{align}
    for all $x\in\mathcal{AR}_i$ where $p\in\left\{1,\dots,h\right\}$, $h=\mathbf{card}(\mathcal{I}(x))$ the number of successive projections of $u_d(x)$ associated with the position $x$, $\mathcal{I}(x)\subset\mathbb{I}$ is the ordered list of the selected obstacles for the successive projections at a position $x\in\mathcal{BL}$,  \textcolor{red}{$\mathcal{AR}_i:=\left\{q\in\mathcal{BL}|i\in\mathcal{I}(q)\right\}$} is the active shadow region (Fig. \ref{fig:fig8}), $u_0(x)=u_d(x)$, $\bar{u}_i=\frac{\sin(\theta_i)}{\sin(\beta_i)}V_p-\frac{\sin(\theta_i-\beta_i)}{\sin(\beta_i)}V_{ci}$, $V_p=\frac{u_{p-1}}{\|u_{p-1}\|}$ and $\beta_i=\angle(V_p,V_{ci})$. Finally, the proposed control law is given by
\begin{align}\label{36}
   u(x)= \begin{cases}
      u_d(x) & \text{if}\;x\in\mathcal{VI},\\
      %\left(\prod\limits_{i\in\mathcal{I}(x)}\pi^{||}(v_i)\right)u_d(x) &
      \mathop{\bigcirc}\limits_{p=0}^h u_p(x)&
      \text{if}\;x\in \mathcal{BL}.
    \end{cases} 
\end{align}
\end{comment}
%%%%%%%%%%%%%%%%%%%%%%%%%%%%%%%%%%%%%%%%%%%%%%%%%%%%%%%%%%%%%%%%%%%%%%%%%%%%%%%%%
%%%%%%%%%%%%%%%%%%%%%%%%%%%%%%%%%%%%%%%%%%%%%%%%%%%%%%%%%%%%%%%%%%%%%%%%%%%%
\section{Safety and Stability Analysis}
In this section, we analyse the safety and stability of the trajectories of the closed-loop system \eqref{12}-\eqref{36}.
Nagumo's theorem (\cite{Nagumo,Set-Theoretic_Methods_in_Control}), offers an important tool to prove safety. One of the statements of this theorem is the one based on Bouligand's tangent cones. 
\begin{definition}Given a closed set $\mathcal{K}$, the tangent cone to $\mathcal{K}$ at $x$ is
$
    \mathcal{T}_{\mathcal{K}}(x):=\{z:\lim_{\tau\rightarrow0}\mathrm{inf}\,\tau^{-1}d(x+\tau z, \mathcal{K})=0\}.
$
\end{definition}
In our case, when $x\in\mathring{\mathcal{F}}$, the tangent cone is the Euclidean space ($\mathcal{T}_{\mathcal{F}}(x)=\mathbb{R}^n$), and since the free space is a sphere world (smooth boundary), the tangent cone at its boundary is a half-space. Nagumo's theorem guarantees, in a navigation problem, that the robot stays inside the free space $\mathcal{F}$. For this to be satisfied, the velocity vector $u(x)$ must point inside (or is tangent to) the free space \cite{souVeloCones}. 
% The uniqueness of the solution of the closed-loop system \eqref{12}-\eqref{36} is proved in the following lemma.
% \begin{lemma}\label{lem2}
% The solution of the closed-loop system \eqref{12}-\eqref{36} is unique.
% \end{lemma}
% \begin{proof}
% See Appendix \ref{appendix:Lemma 2}.
% \end{proof}
In what follows, we rely on Nagumo's theorem to prove the safety of the trajectories generated by our closed-loop system.
\begin{theorem}[Safety]\label{the1}
Consider the closed set $\mathcal{F}$ described in \eqref{8}
and the kinematic system \eqref{12} under the control law \eqref{36}. Then, the closed-loop system \eqref{12}-\eqref{36} admits a unique solution for all $t\geq0$, and the set $\mathcal{F}$ is forward invariant.
\end{theorem}
\begin{proof}
See Appendix \ref{appendix:Theorem 1}.
\end{proof}
%%%%%%%%%%%%%%%%%%%%%%%%%%%%%%%%%%%%%%%%%%%%%%%%%%%%%%
We define the central half-line associated to obstacle $i$
\begin{align}\label{m22}
     \mathcal{L}_i:=\left\{q\in\mathcal{AR}_i|\;\; \beta_i(u_{k-1},q)=0,k=\iota_q^{-1}(i)\right\}.
\end{align}
Let us look for the equilibria of the closed-loop system \eqref{12}-\eqref{36} by setting $u(x)=0$ in \eqref{36}. From the first equation of \eqref{36}, the equilibrium point is $x_d$. From \eqref{eq:recursive-control}, we can rewrite the control at step $p\in\{1,\dots,h(x)\}$ and position $x\in\mathcal{AR}_{\iota_x(p)}$, as $u_p=\sin(\beta_i)\sin^{-1}(\theta_i)\|u_{p-1}\|\bar\xi$\footnote{For simplicity, we drop the arguments $(x,u)$ for the angles $\beta_i$ and $\theta_i$ whenever
clear from context.} where $\iota_x(p)=i$. If we assume that $u_{p-1}\neq0$, and since $\bar{\xi}\in\mathbb{S}^{n-1}$, $u_p=0$ if and only if $\beta_i=0$. Therefore, $u(x)=0$ if $x\in\mathcal{L}_i$, $i\in\mathbb{I}$. Finally, one can conclude that the set of equilibrium points of the system \eqref{12}-\eqref{36} is given by
$
     \zeta:=\{x_d\}\cup(\cup_{i=1}^{m}\mathcal{L}_i).
$
% \begin{comment}
% \begin{figure}[h!]
% \centering
% \includegraphics[scale=0.4]{images/eqpntstac1.png}
% \caption{The left and right figures show the equilibrium points in two-dimensional and three-dimensional sphere worlds.}
% \label{fig:eqpntstac1}
% \end{figure}
% \end{comment}
Now, to ensure that the undesired equilibria $\mathcal{L}_i$ have a repellency property,  we assume the following.
\begin{assumption}\label{as:3}
 For any $i,k\in\mathbb{I}$, $i\neq k$, $\mathcal{L}_k\cap\mathcal{O}_i=\varnothing$. 
\end{assumption}
Assumption 3 restricts the obstacles' configurations in the workspace where no obstacle can intersect the central half-line of another obstacle. However, this will prevent the robot from getting trapped in a central half-line $\mathcal{L}_k$ when avoiding an obstacle $\mathcal{O}_i$.\\
%%% We removed lemma 4 and 5 and included them in theorem 2
\begin{comment}
In the following lemmas, we determine the nature of the equilibria.
\begin{lemma}\label{lem4}
The set of equilibrium points $\mathcal{L}_i$ of the closed-loop system \eqref{12}-\eqref{36} is unstable and a repeller for all $i\in\mathbb{I}$.
\end{lemma}
\begin{proof}
See Appendix \ref{appendix:Lemma 4}.
\end{proof}
\begin{lemma}\label{lem5}
The equilibrium point $x_d$ of the closed-loop system \eqref{12}-\eqref{36} is locally exponentially stable on $\mathcal{F}$ and attractive for $x(0)\in\mathcal{F}\setminus\bigcup\limits_{i=1}^{m}\mathcal{L}_i$.
\end{lemma}
\begin{proof}
see Appendix \ref{appendix:Lemma 5}.
\end{proof}
\end{comment}
In what follows, we present our main theorem:
\begin{theorem}\label{the2}
Consider the free space $\mathcal{F}\subset\mathbb{R}^n$ described in \eqref{8} and the closed-loop system \eqref{12}-\eqref{36}. Under Assumptions \ref{as:1}, \ref{as:2} and \ref{as:3}, the following statements hold:
\begin{itemize}
\item All trajectories converge to the set $ \zeta=\{x_d\}\cup(\cup_{i=1}^{m}\mathcal{L}_i)$.
\item The set of equilibrium points $\cup_{i=1}^{m}\mathcal{L}_i$ is unstable and a repeller.
\item The equilibrium point $x_d$ is locally exponentially stable on $\mathcal{F}$ and attractive from all $x(0)\in\mathcal{F}\setminus\cup_{i=1}^{m}\mathcal{L}_i$.
\item From any initial position $x_0\in\mathcal{F}\setminus\cup_{i=1}^{m}\mathcal{L}_i$, the trajectory $x(t)$ generates a quasi-optimal obstacle avoidance maneuver.
\end{itemize}
\end{theorem}
\begin{proof}
See Appendix \ref{appendix:the2}.
\end{proof}
Theorem \ref{the2} shows that the desired equilibrium point $x_d$ is almost globally asymptotically stable (since $\cup_{i=1}^{m}\mathcal{L}_i$ is Lebesgue measure zero) and that all trajectories of the closed-loop system are safe and quasi-optimal, in the sense of Definition \ref{def2}. In the next section, we illustrate this optimality property in different scenarios.
%%%%%%%%%%%%%%%%%%%%%%%%%%%%%%%%%%%%%%%%%%%%%%%%%%%%%%%%%%
\section{Numerical simulation}
\begin{figure}[h!]
     \centering
     \subfloat[]{\includegraphics[width=0.4\linewidth,height=0.4\linewidth,keepaspectratio]{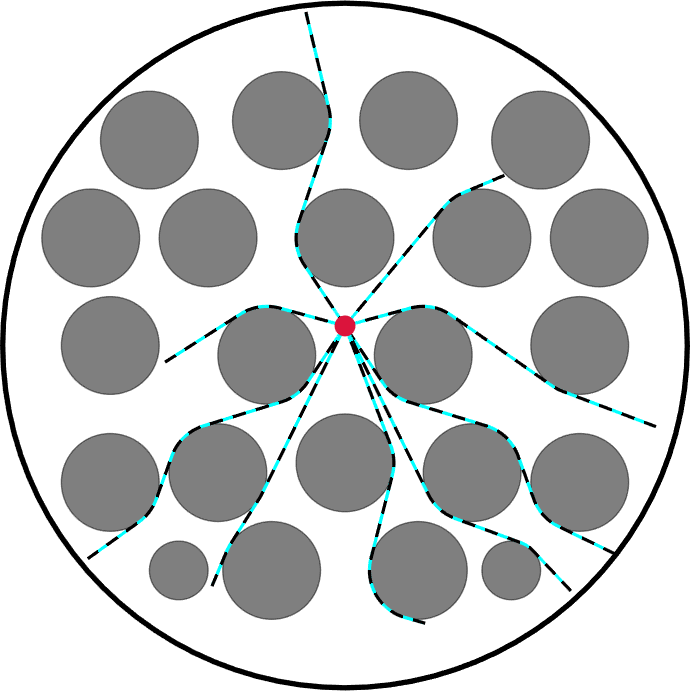}\label{s2}}
    % \hspace{0.01cm}
     \subfloat[]{\includegraphics[width=0.4\linewidth,height=0.4\linewidth,keepaspectratio]{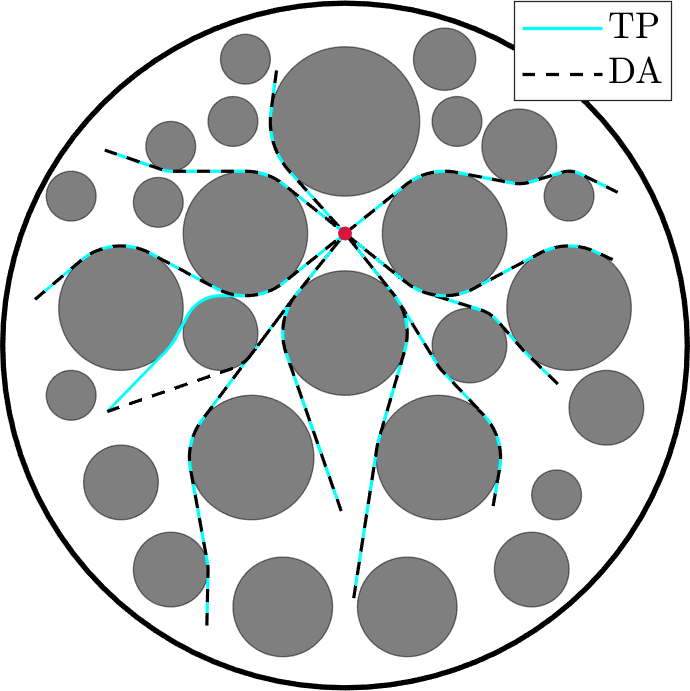}\label{s3}}\\
     \subfloat[]{\includegraphics[width=0.4\linewidth,height=0.4\linewidth,keepaspectratio]{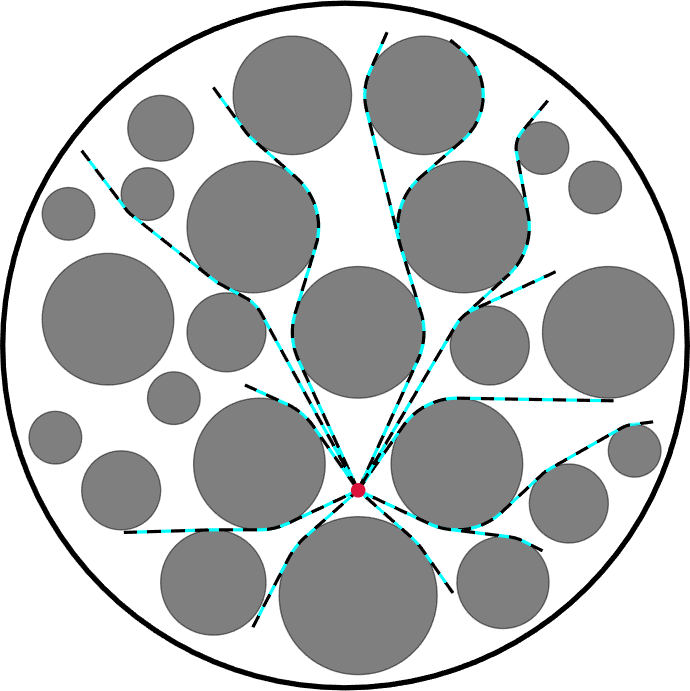}\label{s4}}
     %\hspace{0.01cm}
     \subfloat[]{\includegraphics[width=0.4\linewidth,height=0.4\linewidth,keepaspectratio]{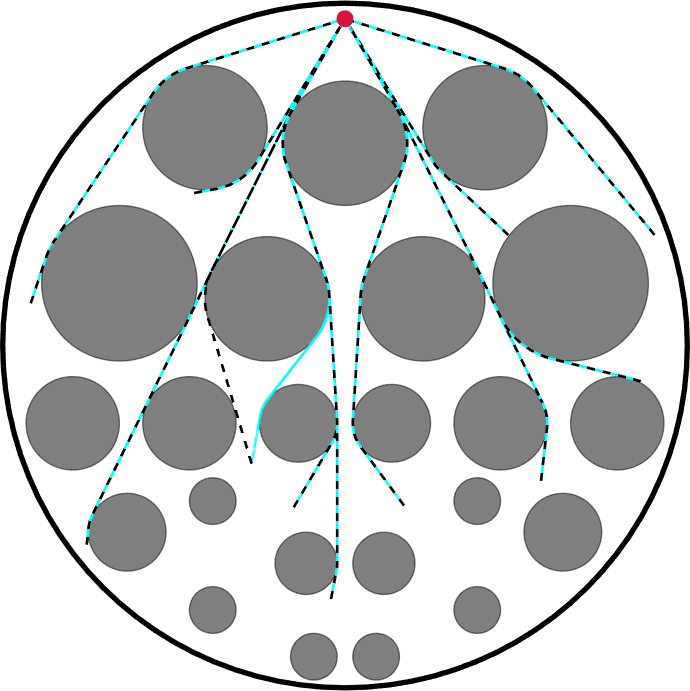}\label{s5}}
     \caption{Some samples among 100 tests between (DA) and our approach (TP) in 4 different spaces.}
     \label{4spaces}
\end{figure}

\begin{figure}[h!]
     \centering
     \subfloat[]{\includegraphics[width=0.4\linewidth,height=0.4\linewidth,keepaspectratio]{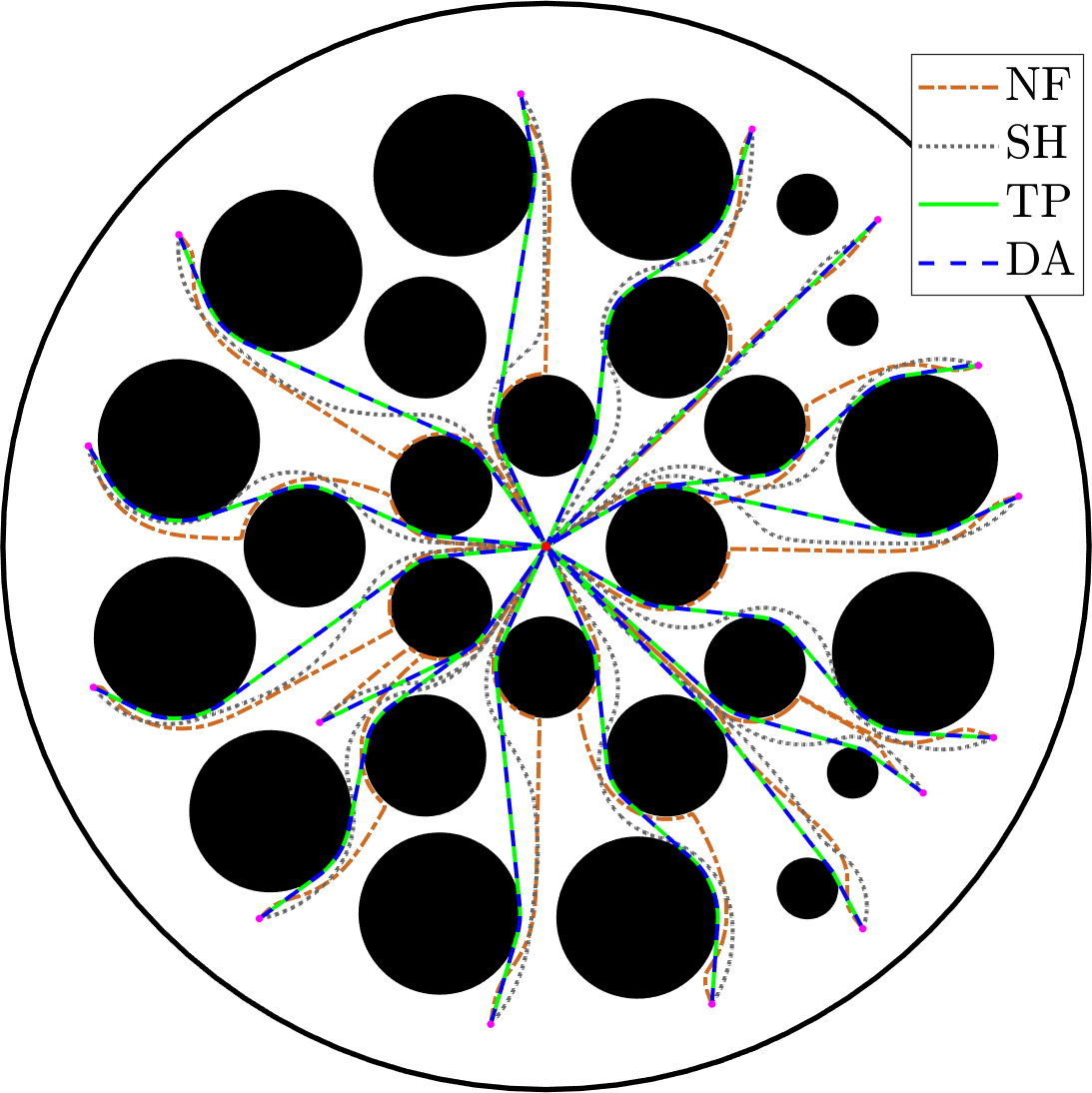}\label{ma2D}}
     \subfloat[]{\includegraphics[width=0.4\linewidth,height=0.4\linewidth,keepaspectratio]{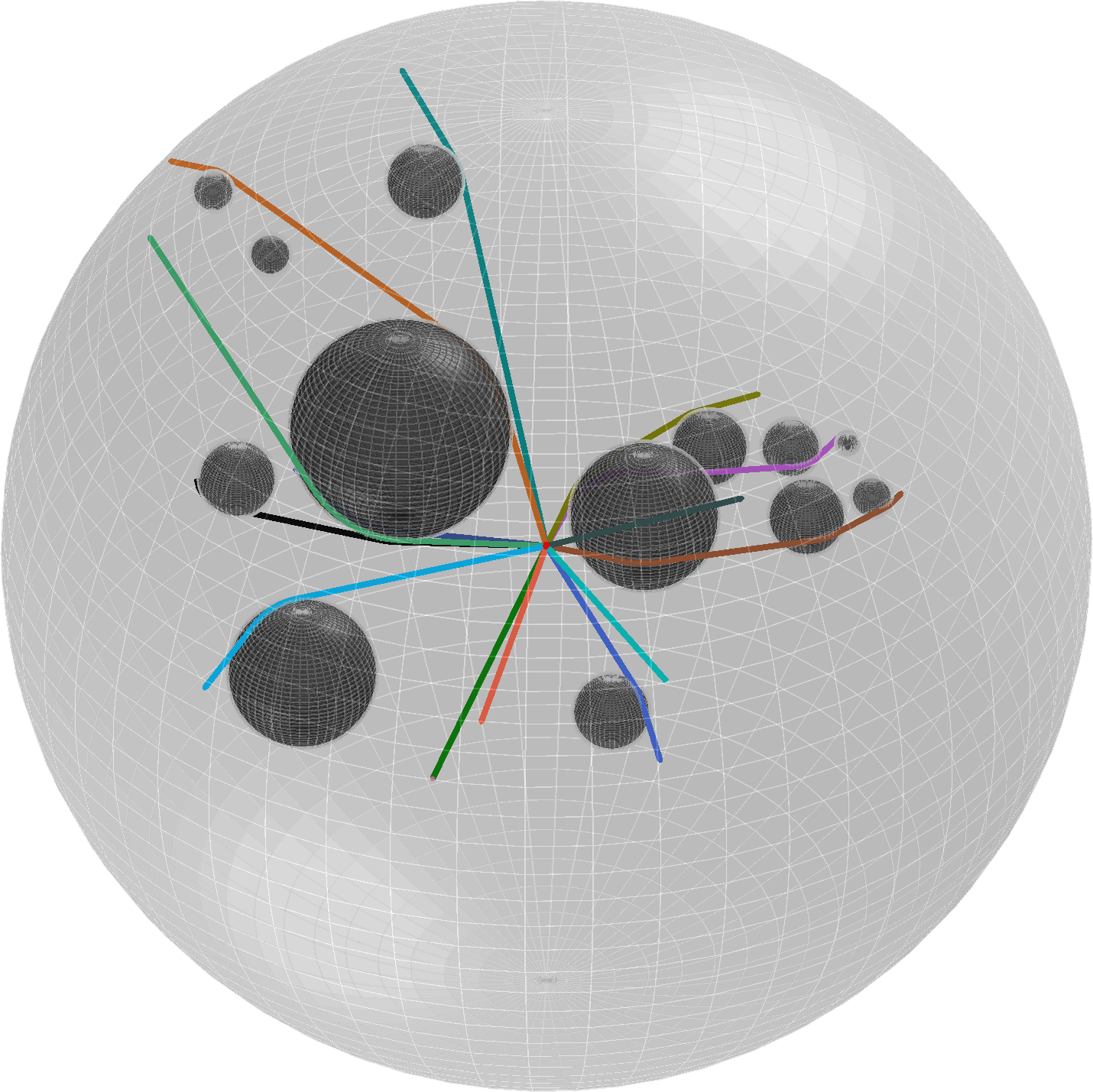}\label{mb3D}}
     \caption{Robot safe navigation from fifteen different initial positions. Fig. (a) is a comparison between our method TP, SH, NF, and DA in a two-dimensional sphere world, while Fig (b) only shows the performance of our approach in a three-dimensional sphere world.}
     \label{simmultiobs}
\end{figure}

To explore the extent of what our quasi-optimal avoidance maneuver can offer in terms of the shortest path in the multiple obstacle case, we compare the trajectories of our method (TP) to the shortest paths obtained with Dijkstra's algorithm (DA) on a visibility tangent graph in five different and highly congested two-dimensional spaces where the first space is represented in Fig. \ref{ma2D} and the four other spaces are represented in Fig. \ref{4spaces}. In each space, we take 100 random initial conditions, and we count the number of perfect matching of the paths. The summarized results in Table \ref{table_1} show a high rate of success, while the failures of taking the shortest path can be explained through the fact that, at each instant, our approach considers the set $\mathcal{I}(x)$ engaged in the nested successive projections that may lead to a non-optimal path. Compared to Dijkstra's algorithm, which takes the shortest path from the visibility tangent graph that considers all the obstacles.\\
To visualize the properties of our approach, we consider thirteen obstacles in two different scenarios. In the first scenario, we assume that the robot evolves in $\mathbb{R}^2$ where the destination $x_d=[0\;0]^\top$, while the second scenario is in $\mathbb{R}^3$ and the goal is $x_d=[0\;0\;0]^\top$. In both cases, we consider fifteen different initial positions. A comparison of our approach with the navigation function approach (NF) \cite{k_R_90} and the separating hyperplane approach (SH) \cite{Arslan2019} is established in the two-dimensional space. The simulation results in Fig. \ref{simmultiobs} show that all the trajectories generated by our control are safe and converge to the red target. In addition, Fig. \ref{ma2D} shows the superiority of our approach over the two other methods in terms of the length of the generated collision-free paths, where it generates the same paths as DA.
\begin{table}[h!]
\caption{Success rate of the perfect match between the paths generated by our control and the paths of DA.}
\label{table_1}
\begin{center}
\begin{tabular}{|c||c||c||c||c|}
\hline
Space 1 & Space 2 & Space 3 & Space 4 & Space 5\\
\hline
96$\%$ & 98$\%$ &93$\%$ &97$\%$ &97$\%$\\
\hline
\end{tabular}
\end{center}
\end{table}
%%%%%%%%%%%%%%%%%%%%%%%%%%%%%%%%%%%%%%%%%%%%%%%%%%%%%%%%%%%%%%%%%%%%%%%%%%%%%%%%
\section{Conclusion}
\begin{comment}
We proposed a quasi-optimal continuous feedback control strategy, with almost global asymptotic stability guarantees, for the autonomous navigation problem in an n-dimensional sphere world. The proposed strategy consists in steering the robot tangentially to the blocking obstacles through successive projections of the nominal control on the obstacles enclosing cones. As a consequence, the intermediary obstacle avoidance maneuvers are optimal resulting in a quasi-optimal overall collision-free path. \textcolor{red}{Another interesting byproduct of the proposed control strategy is that it generates, in some cases, less undesired equilibria than the number of obstacles, a nonexistent feature in the available feedback based autonomous navigation schemes.}\footnote{\color{red} Assumption 3 excludes theses configurations unfortunately} 
Extending the proposed approach to arbitrarily shaped obstacles, with global asymptotic stability guarantees, is an interesting problem that will be the topic of pour future investigations.
\end{comment}
We have proposed a quasi-optimal continuous feedback control strategy, with almost global asymptotic stability guarantees, for the autonomous navigation problem in an $n$-dimensional sphere world. The proposed strategy consists in steering the robot tangentially to the blocking obstacles through successive projections of the nominal control onto the obstacles enclosing cones. Consequently, the intermediary obstacle avoidance maneuvers are optimal, resulting in a quasi-optimal overall collision-free path. We recognize that the price to pay to obtain the claimed results in the paper is a somewhat restrictive assumption on the obstacles configuration (Assumption 3) that needs to be relaxed in our future investigations. Extending the proposed approach to arbitrarily shaped obstacles, with global asymptotic stability guarantees, is another interesting problem that will be the main focus of our future work.
%%%%%%%%%%%%%%%%%%%%%%%%%%%%%%%%%%%%%%%%%%%%%%%%%%%%%%%%%%%%%%%%%%%%%%%%%%%%%%%%
\section*{APPENDIX}
\subsection{Proof of Lemma 1}\label{appendix:Lemma 1}
Minimizing the angle $\angle(x_d-x,v_i)$ is equivalent to minimizing the cost function $g(v_i)=1-V_d^\top\frac{v_i}{\|v_i\|}$ with $V_d=(x_d-x)/\|x_d-x\|$ under the constraint $\Gamma(v_i)=\frac{v_i^\top V_{ci}}{\|v_i\|}-\cos(\theta_i)=0$ with $V_{ci}=(c_i-x)/\|c_i-x\|$. We define the Lagrangian associated to the optimization problem \eqref{min} by $ L_{\lambda}(v_i)=g(v_i)-\lambda \Gamma(v_i)$ where $\lambda$ is the Lagrange multiplier. The optimum is the solution of $\nabla_{v_i,\lambda}L_{\lambda}(v_i)=0$ which gives
\begin{align}
        \pi^{\bot}(v_i)(V_d+\lambda V_{ci})=0,\quad 
        \frac{v_i^\top V_{ci}}{\|v_i\|}-\cos(\theta_i)=0.\label{32}
\end{align}
From the first equation, we have $v_i=\alpha(V_d+\lambda V_{ci})$ for some $\alpha\in\mathbb{R}$. Substituting this into \eqref{32} and then \label{31}, we can solve for $\lambda$ and find
\begin{align}
    \lambda_{1,2}=-\frac{\sin(\theta_i\pm\beta_i)}{\sin(\theta_i)},%\quad
    %\lambda_2=-\frac{\sin(\theta_i-\beta_i)}{\sin(\theta_i)}\label{lambda2}.
\end{align}
Therefore, we obtain two vectors $v_i^1$ and $v_i^2$ such that
\begin{align}
      v_i^{1,2}=\alpha_{1,2}\left(V_d-\frac{\sin(\theta_i\pm\beta_i)}{\sin(\theta_i)}V_{ci}\right),
      %\\ v_i^2=\alpha_2\left(V_d-\frac{\sin(\theta_i-\beta_i)}{\sin(\theta_i)}V_{ci}\right),
\end{align}
where $\alpha_1\leq0$ and $\alpha_2\geq0$.
The value of $g$ at the two solutions is as follows:
\begin{align*}
  \small
    g(v_i^1)=1+\cos(\theta_i+\beta_i),\;
    g(v_i^2)=1-\cos(\theta_i-\beta_i),
\end{align*}
and $g(v_i^1)-g(v_i^2)=2\cos(\theta_i)\cos(\beta_i)\geq0$ which implies that
{\small
\begin{align}
  \mathcal{U}_1(x)=\{\alpha_2(V_d-\sin^{-1}(\theta_i)\sin(\theta_i-\beta_i)V_{ci})|\; \alpha_2\geq0\}.
\end{align}}
When $x\in\mathcal{S}(x_d,c_i)$, $\theta_i=\beta_i$ and for all $v_i\in\mathcal{U}_1$, $v_i\in\mathcal{U}_2$. Therefore, $\alpha_2V_d=u_d(x)$ which implies that $\alpha_2=\gamma\|x_d-x\|$. One can conclude that the set $\mathcal{U}_1\cap\mathcal{U}_2$ is a singleton and the unique solution is given by
\begin{align*}
\small
    u(x)&=\gamma\|x_d-x\|\left(V_d-\frac{\sin(\theta_i-\beta_i)}{\sin(\theta_i)}V_{ci}\right)=\xi(u_d(x),x,i), 
\end{align*}
where the last equation is obtained after some straightforward manipulation.
% \begin{align}\label{33}
%         v_i=\alpha(V_d+\lambda V_{ci}).
% \end{align}
% where $\alpha\in\mathbb{R}$. We substitute \eqref{33} into \eqref{32} to get
% \begin{align}\label{34}
%         \alpha(\cos(\beta_i)+\lambda)=\|\alpha(V_d+\lambda V_{ci})\|\cos(\theta_i).
% \end{align}
% We square \eqref{34} and get 
% \begin{align}\label{35}
% \sin^2(\theta_i)\lambda^2+2\cos(\beta_i)\sin^2(\theta_i)\lambda+\cos^2(\beta_i)-\cos^2(\theta_i)=0.
% \end{align}
% The solutions of \eqref{35} are
% \begin{align}
%     \lambda_1=-\frac{\sin(\theta_i+\beta_i)}{\sin(\theta_i)},\quad
%     \lambda_2=-\frac{\sin(\theta_i-\beta_i)}{\sin(\theta_i)}\label{lambda2}.
% \end{align}
% We substitute \eqref{lambda2} in \eqref{34} and obtain $-\alpha_1=|\alpha_1|$ and $\alpha_2=|\alpha_2|$ which implies that $\alpha_1\leq0$ and $\alpha_2\geq0$. 
%%%%%%%%%%%%%%%%%%%%%%%%%%%%%%%%%%%%%%%%%%
\subsection{Proof of Lemma 2}\label{appendix:Lemma 6} 
\begin{figure}[h!]
\centering
\includegraphics[scale=0.3]{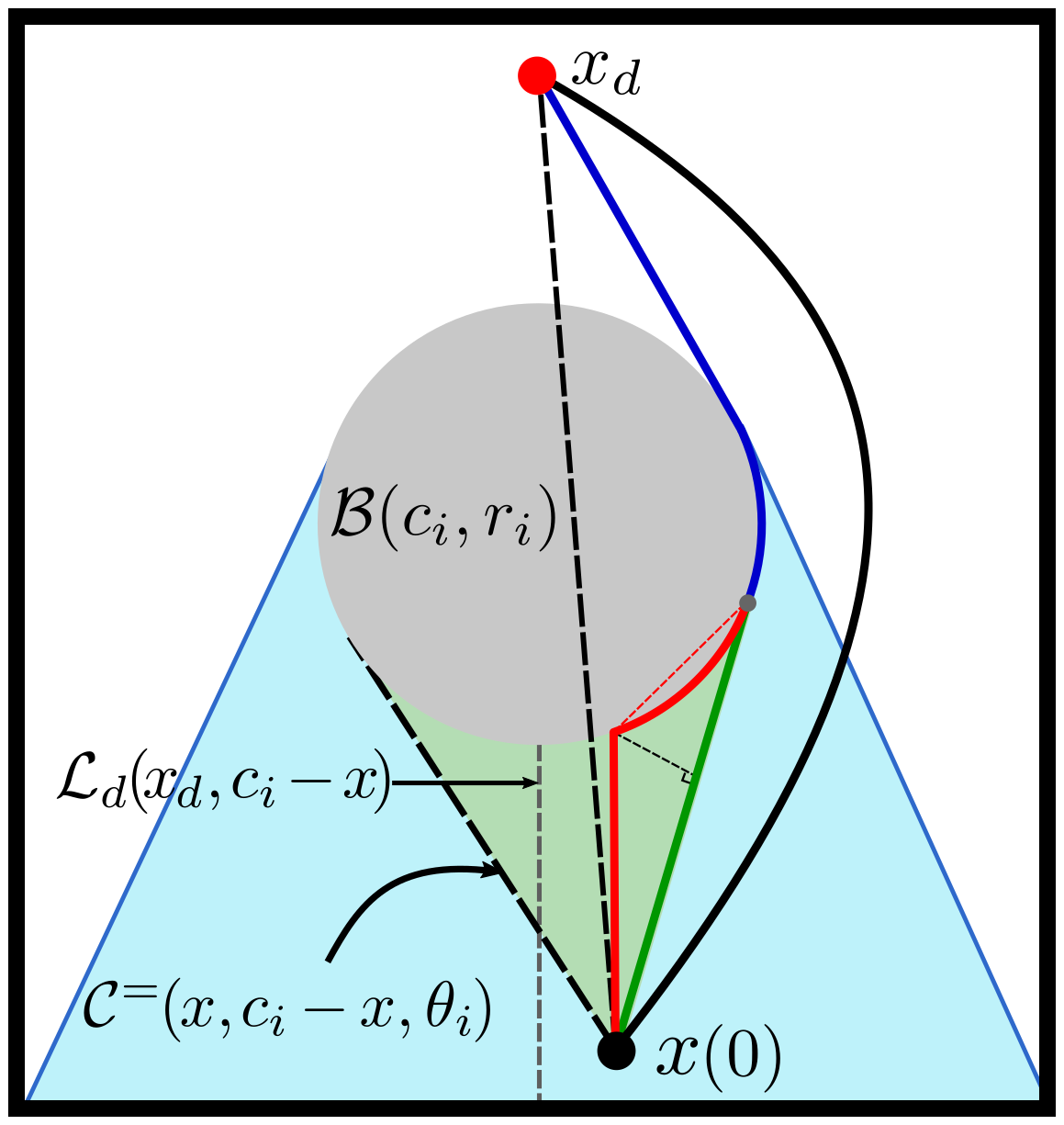}
\caption{Shortest path in a single-obstacle sphere world.}
\label{fig:fig6}
\end{figure}

Let $x(0)\in\mathcal{F}\setminus\mathcal{L}_d(x_d,c_i)$. Then we have two situations. First, when $x(0)\in\mathcal{VL}$, the trajectory $x(t)$ is a line-segment which is the closest path. Now, when $x(0)\in\mathcal{D}(x_d,c_i)$, there are two types of possible trajectories: trajectories inside the enclosing cone $\mathcal{C}^{\leq}_{\mathcal{F}}(x,c_i-x,\theta_i)$ and trajectories outside this cone. We show that the trajectory generated by the closed-loop system \eqref{12}-\eqref{25} on the enclosing cone $\mathcal{C}^{=}_{\mathcal{F}}(x,c_i-x,\theta_i)$ has the minimum length. For the first type of trajectory, we only consider the ones between the line segment $\mathcal{L}_s(x(0), x_d)$ and the closest tangent to it (green segment in Fig. \ref{fig:fig6}) among the cone enclosing the obstacle (the red trajectory in Fig. \ref{fig:fig6} is an example). All these trajectories will merge with our trajectory, which is on the closest tangent (as shown in Lemma \ref{lem1}), at the intersection point of the tangent with the obstacle. Since, before the intersection point, our trajectory is a line segment, we can conclude that it is the shortest path. The best that can be achieved outside the cone for a smooth trajectory is a dilated version of our trajectory (larger radius of curvature) which is longer than ours (black path in Fig. \ref{fig:fig6}).
%%%%%%%%%%%%%%%%%%%%%%%%%%%%%%%%%%%%%%%%%%%%%%%%%%%%%%%%%%%%%%%%%%%%%%%%%%%
\subsection{Proof of Theorem 1}\label{appendix:Theorem 1}
First we prove that the closed-loop system admits a unique solution. The control is Lipschitz on $\mathcal{VI}$ since $u(x)=u_d(x)$ is continuously differentiable. When  $x\in\mathcal{BL}$, for simplicity, we denote $\sin(\theta_{\iota_x(p)}(q)-\beta_{\iota_x(p)}(u_{p-1}(x),q))$ by $s_{\iota_x(p)}^s(q)$ and $\frac{\sin(\beta_{\iota_x(p)}(u_{p-1}(x),q))}{\sin(\theta_{\iota_x(p)}(q))}$ by $s_{\iota_x(p)}^d(q)$ where $p\in\{1,\dots,h(x)\}$. After manipulation, the control \eqref{36} can be expressed as $u(x)=u_d(x)-\gamma\|x-x_d\|\sum\limits_{p=1}^{h(x)}\prod\limits_{k=1}^{p-1} s_{\iota(k)}^d(x)\frac{s_{\iota(p)}^s(x)}{r_{\iota(p)}}(c_{\iota(p)}-x)$, and we prove that it is one-sided Lipschitz as follows:
    \begin{align*}
    \begin{split}
        &\left(u(x)-u(y)\right)^\top(x-y)=-\gamma\|x-y\|^2\\
        &-\gamma\|x_d-x\|\sum\limits_{p=1}^{h(x)}\prod\limits_{k=1}^{p-1} s_{\iota(k)}^d(x)\frac{s_{\iota(p)}^s(x)}{r_{\iota(p)}}(c_{\iota(p)}-x)^\top(x-y)\\&+\gamma\|x_d-y\|\sum\limits_{p=1}^{h(y)}\prod\limits_{k=1}^{p-1} s_{\iota(k)}^d(y)\frac{s_{\iota(p)}^s(y)}{r_{\iota(p)}}(c_{\iota(p)}-y)^\top(x-y)\\
      %\end{split}
    %\end{align*}   
    %\begin{align*}
    %\begin{split}
        %&\left(u(x)-u(y)\right)^\top(x-y)
        &\leq-\gamma\|x-y\|^2\\
        &+\gamma\|x_d-x\|\|x-y\|\sum\limits_{p=1}^{h(x)}\prod\limits_{k=1}^{p-1} s_{\iota(k)}^d(x)\frac{s_{\iota(p)}^s(x)}{r_{\iota(p)}}\|c_{\iota(p)}-x\|\\&+\gamma\|x_d-y\|\|x-y\|\sum\limits_{p=1}^{h(y)}\prod\limits_{k=1}^{p-1} s_{\iota(k)}^d(y)\frac{s_{\iota(p)}^s(y)}{r_{\iota(p)}}\|c_{\iota(p)}-y\|.
    \end{split}
    \end{align*}
    Note that $\forall x\in\mathcal{BL}$ and $\forall p\in\{1,\dots,h(x)\}$, $0\leq s_{\iota(p)}^d(x)\leq1$, $0\leq s_{\iota(p)}^s(x)\leq1$, $\|c_{\iota(p)}-x\|\leq 2r_0-r_{\iota(p)}$ and $\|x_d-x\|\leq 2r_0$, which implies that there exists $M>0$ such that $\|x_d-x\|\sum_{p=1}^{h(x)}\frac{\|c_{\iota(p)}-x\|}{r_{\iota(p)}}\leq M\|x-y\|$. Therefore,
    \begin{align*}
    \begin{split}
        &\left(u(x)-u(y)\right)^\top(x-y)\leq-\gamma\|x-y\|^2+\\&\qquad\gamma M_1\|x-y\|^2+\gamma M_2\|x-y\|^2\\&\quad\leq\gamma(-1+M_1+M_2)\|x-y\|^2\leq L\|x-y\|^2.
    \end{split}
    \end{align*}
One can take $L=\gamma(-1+M_1+M_2)$ where $M_1>0$, $M_2>0$ and $M_1+M_2>1$. The control \eqref{36} is one-sided Lipschitz \cite{uniqueness} when $x\in\mathcal{BL}($ and is Lipschitz when $x\in\mathcal{VI}$. Thus, according to \cite[Proposition 2]{uniqueness}, the closed-loop system \eqref{12}-\eqref{36} has a unique solution for all $x(0)\in\mathcal{F}$.
% \begin{figure}[h!]
%     \centering
%     \includegraphics[scale=0.35]{images/fig9.png}
%     \caption{Illustration of Bouligand's tangent cones (pink shapes) in a two-dimensional sphere world. The red arrows show the adherence of the control \eqref{36} to Nagumo's condition.}
%     \label{fig:fig9}
%     \end{figure}
Now, we prove forward invariance using Nagumo's theorem. We only need to verify Nagumo's condition at the free space boundary as it is trivially met when $x\in\mathring{\mathcal{F}}$ where $\mathcal{T}_{\mathcal{F}}(x)=\mathbb{R}^n$. Since the free space is a sphere world, the tangent cone on its boundary is the half-space $\mathcal{C}^{\leq}_{\mathbb{R}^n}(x,-x,\frac{\pi}{2})$ when $x\in\partial\mathcal{W}$ and $\mathcal{C}^{\geq}_{\mathbb{R}^n}(x,c_i-x,\frac{\pi}{2})$ when $x\in\partial\mathcal{O}_i$. We consider an obstacle $\mathcal{O}_i$  and verify Nagumo's condition in three regions of the free space.\\
In the first region, When $x\in\partial\mathcal{W}$, $\mathcal{T}_{\mathcal{F}}(x)=\mathcal{C}^{\leq}_{\mathbb{R}^n}(x,-x,\frac{\pi}{2})$ and two sub-regions must be considered.
            \begin{itemize}
             \item $x\in\partial\mathcal{W}\cap\partial\mathcal{BL}$: Since $u(x)\in \mathcal{C}^=_{\mathcal{F}}(x,c_i-x,\theta_i)$ and $\mathcal{C}^=_{\mathcal{F}}(x,c-x,\theta)\subseteq\mathcal{C}^{\leq}_{\mathbb{R}^n}(x,-x,\frac{\pi}{2})$, we conclude that $u(x)\in\mathcal{T}_{\mathcal{F}}(x)$.
             \item $x\in\partial\mathcal{W}\setminus\partial\mathcal{BL}$: Since $u_d(x)\in\mathcal{F}$ and $\mathcal{F}\subseteq\mathcal{C}^{\leq}_{\mathbb{R}^n}(x,-x,\frac{\pi}{2})$, we conclude that $u(x)=u_d(x)\in\mathcal{T}_{\mathcal{F}}(x)$.
            \end{itemize}
In the second region, $x\in\partial\mathcal{O}_i\cap\mathcal{AR}_i$ and $\mathcal{T}_{\mathcal{F}}(x)=\mathcal{C}^{\geq}_{\mathbb{R}^n}(x,c_i-x,\frac{\pi}{2})$. Since $u(x)\in\mathcal{C}^{=}_{\mathbb{R}^n}(x,c_i-x,\frac{\pi}{2})\subset\mathcal{C}^{\geq}_{\mathbb{R}^n}(x,c_i-x,\frac{\pi}{2})$, one concludes that $u(x)\in\mathcal{T}_{\mathcal{F}}(x)$.
Finally, in the last region, $x\in\partial\mathcal{O}_i\setminus\mathcal{AR}_i$ and $\mathcal{T}_{\mathcal{F}}(x)=\mathcal{C}^{\geq}_{\mathbb{R}^n}(x,c_i-x,\frac{\pi}{2})$. Since $x\notin\mathcal{AR}_i$, $\forall p\in\{0,\cdots,h(x)\}$, obstacle $\mathcal{O}_i$ is not selected in the successive projections $(\iota_x(p)\neq i)$ and $u_p(x)\notin\mathcal{C}^{\leq}_{\mathbb{R}^n}(x,c_i-x,\frac{\pi}{2})$. Therefore, $u(x)$ must be in the complement of the enclosing cone to the obstacle $\mathcal{O}_i$. Thus, one can conclude that $u(x)\in\mathcal{C}^{\geq}_{\mathbb{R}^n}(x,c_i-x,\frac{\pi}{2})=\mathcal{T}_{\mathcal{F}}(x)$. 

Since $\forall x\in\mathcal{F},\;\; u(x)\in\mathcal{T}_{\mathcal{F}}(x)$ 
and the solution of the closed-loop system \eqref{12}-\eqref{36} is unique, it follows that the free space $\mathcal{F}$ is positively invariant and the closed-loop system \eqref{12}-\eqref{36} is safe.
\subsection{Proof of Theorem 2}\label{appendix:the2}
Four points must be demonstrated, safety of the system, quasi-optimality of the avoidance maneuver, stability of the destination $x_d$ and instability of the remaining equilibria.\\
\begin{comment}
\begin{enumerate}
    \item Safety: by virtue of Theorem \ref{the1}, the closed-loop system \eqref{12}-\eqref{36} is safe for all $x\in\mathcal{F}$.
    \item Stability of $x_d$: according to Lemma \ref{lem5}, the destination $x_d$ is locally asymptotically stable and almost globally attractive equilibrium point on $\mathcal{F}$.
    \item Instability of $\mathcal{L}_i$ with $i\in\mathbb{I}$: Lemma \ref{lem4} states that $\Bar{x}_i\in\mathcal{L}_i$ is an unstable equilibrium point and a repeller.
    \item Quasi-optimal obstacles avoidance maneuver: Since the control input \eqref{36} is a composition of the projection from Lemma \ref{lem1}, which generates, according to Lemma \ref{lem6}, the shortest path for a considered obstacle. Therefore, the trajectory $x(t)$ of the closed-loop system \eqref{12}-\eqref{36} generates quasi-optimal obstacles avoidance maneuver as per Definition \ref{def2}.
\end{enumerate}
\end{comment}
First, we start by proving the safety of our system. By virtue of Theorem \ref{the1},  the closed-loop system \eqref{12}-\eqref{36} is safe for all $x\in\mathcal{F}$.\\
Next, we prove the instability of the equilibria $\mathcal{L}_i$ with $i\in\mathbb{I}$. Consider an obstacle $\mathcal{O}_i$ and the set of its ancestors $\mathcal{AN}_i\subset\{0,\dots,m\}$ where $i\notin\mathcal{AN}_i$ and the element $0$ refers to the destination. Each ancestor generates a subset $\mathcal{AR}_i^{k}$ of the active shadow region $\mathcal{AR}_i$ where  $\mathcal{AR}_i^{k}\cap\mathcal{AR}_i^j=\varnothing$ for all $k,j\in\mathcal{AN}_i$ and $k\neq j$, and $\mathcal{AR}_i=\bigcup\limits_{k\in\mathcal{AN}_i}\mathcal{AR}_i^{k}$. The subset generated by the ancestor responsible for the creation of the central half-line $\mathcal{L}_i$ is denoted by $\mathcal{AR}_i^*$. We consider the set of equilibrium points $\mathcal{L}_i$ where $i\in\mathbb{I}$ and we define the extended central half-line $\mathcal{L}_i^e:=\left\{q\in\mathcal{W}|q=c_i+\delta\frac{y-c_i}{\|y-c_i\|}\,, y\in\mathcal{L}_i\,,\delta\in\mathbb{R}_{>0}\right\}$. We also define the cylinder inside $\mathcal{AR}_i^*$ by $\mathcal{CY}_i(e_i,\epsilon_i):=\left\{q\in\mathcal{AR}_i^*|\,d\left(q,\mathcal{L}_i^e\right)<e_i\,\text{and}\,\|q-c_i\|<\epsilon_i,\,e_i,\epsilon_i\in\mathbb{R}_{>0}\right\}$, where $\epsilon_i$ is such that for all $x\in\mathcal{CY}_i(e_i,\epsilon_i)$, the considered obstacle $\mathcal{O}_i$ is the last on the successive projection, {\it i.e.,} $\forall x\in\mathcal{CY}_i(e_i,\epsilon_i),\,\iota_x(h(x))=i$.  We also define the set $U:=\mathcal{CY}_i(e_i,\epsilon_i)\setminus  \mathcal{L}_i$. Let the equilibrium point $\bar{x}_i\in\mathcal{L}_i$ and $V_1(x)=1-\frac{(\Bar{x}_i-c_i)^\top}{r_i}\frac{(x-c_i)}{||x-c_i||}$ where $V_1(\Bar{x}_i)=0$ and $V_1(x)>0$ for all $x\in U$.
\begin{align*}
    \Dot{V}_1(x)&=\frac{\partial V_1(x)}{\partial x}^\top\Dot{x}=-\frac{(\Bar{x}_i-c_i)^\top}{r_i}J_x\left(\frac{(x-c_i)}{||x-c_i||}\right)u(x)\\
    &=-K\Bar{V}^\top_{ci}\pi^{\bot}(V_{ci})\Bar{\xi},
\end{align*}
where $K=\frac{\|u(x)\|\|\Bar{x}_i-c_i\|}{r_i||x-c_i||}>0$, $V_{ci}=\frac{(c_i-x)}{\|c_i-x\|}$, $\bar{V}_{ci}=\frac{(\Bar{x}_i-c_i)}{\|\Bar{x}_i-c_i\|}$ and $\bar{\xi}=\frac{\sin(\theta_i)u_{h(x)-1}}{\sin(\beta_i)\|u_{h(x)-1}\|}-\frac{\sin(\theta_i-\beta_i)}{\sin(\beta_i)}V_{ci}$. Recall that the control, at position $x\in\mathcal{BL}$, is written as $u(x)=u_{h(x)}(x)=\sin(\beta_i)\sin^{-1}(\theta_i)\|u_{h(x)-1}\|\bar\xi$ with $i=\iota_x(h(x))$. Then, the vectors $u_{h(x)}$, $u_{h(x)-1}$ and $V_{ci}$ are on the same $2$D plane. Moreover, the central half-line $\mathcal{L}_{i}$ is generated by the overlap between $V_{ci}$ and $u_{h(x)-1}/\|u_{h(x)-1}\|$ when $x\in\mathcal{AR}_{i}^*$. Therefore,  the vectors $u_p(x)$, $u_{h(x)-1}(x)$, $V_{ci}$ and the central half-line $\mathcal{L}_i$ are on the same plane. Similarly, the vectors $\bar{\xi}$, $V_{ci}$ and $\Bar{V}_{ci}$ are on same plane, and thus, 
\begin{align*}
    \Dot{V}_1(x)&=-K\left(\cos(\sigma_i)-\cos(\theta_i)\cos(\sigma_i+\theta_i) \right),\\
    &=K\left(\cos(\sigma_i)\sin^2(\theta_i)+\cos(\theta_i)\sin(\sigma_i)\sin(\theta_i) \right)\\
    &=K\sin(\theta_i)\sin(\sigma_i-\theta_i),
\end{align*}
where $\frac{\pi}{2}<\angle(\Bar{V}_{ci},\Bar{\xi})=\sigma_i\leq\pi$ and $0<\sigma_i-\theta_i<\pi$.\\
$\Dot{V_1}(\Bar{x}_i)=0$ and $\Dot{V_1}(x)>0$ for all $x\in U$. According to Chetaev's theorem \cite[Theorem 4.3]{Khalil}, the equilibrium point $\Bar{x}_i$ is unstable, and for any $x_0\in U$, $x(t)$ must leave $U$ from all directions else than the surface of the obstacle (due to the safety of the system). Moreover, considering assumption \ref{as:3} and that the control $u(x)$ is tangent to the last obstacle $\iota(h(x))$, we can conclude that $x(t)$ must leave the active shadow region of the last obstacle. Therefore, any equilibrium $\Bar{x}_i\in\mathcal{L}_i$ is unstable and a repeller, and the blind set has the following property:
\begin{align}\label{prop}
    \exists t^*\in\mathbb{R}_{>0},\;\forall x(0)\in\mathcal{BL}\setminus\bigcup\limits_{i=1}^m\mathcal{L}_i,\;x(t^*)\notin\mathcal{BL}.
\end{align}
% \begin{figure}[h!]
%     \centering
%     \includegraphics[scale=0.32]{images/proof_lemm_unst.png}
%     \caption{Illustration of the active shadow regions.}
%     \label{fig:proof_unst}
% \end{figure}
Now we prove the stability of the $x_d$. We consider the equilibrium point $x_d$ and the positive definite function $V_2(x)=\frac{1}{2}||x-x_d||^2$. The closed-loop system \eqref{12}-\eqref{36} reduces to $\Dot{x}=-\gamma(x-x_d)$ When $x\in\mathcal{VI}$. Therefore, $\Dot{V}_2(x)= -\gamma||x-x_d||^2<0$. We can conclude that the destination is locally exponentially stable and almost globally asymptotically stable using the property \eqref{prop}.
% \begin{comment}
% \footnotesize
% \begin{align}
%     \Dot{V}_2(x)&=\frac{\partial V_2(x)}{\partial x}^\top\Dot{x},\\
%     &=(x-x_d)^\top u(x),\\
%     &=\begin{cases}
%       -\gamma||x-x_d||^2 &\text{if}\;x\in\mathcal{VI},\\
%       -u_d(x)^\top\prod\limits_{p=1}^{h(x)}\frac{\sin(\beta_{\iota(p)})\pi^{||}(\bar{u}_{\iota(p)})}{\cos(\theta_{\iota(p)}-\beta_{\iota(p)})\sin(\theta_{\iota(p)})}u_d(x) &\text{if}\;x\in\mathcal{BL},
%       \end{cases} 
% \end{align}
% \normalsize
% where $\bar{u}_i=\frac{\sin(\theta_i)}{\sin(\beta_i)}V_p-\frac{\sin(\theta_i-\beta_i)}{\sin(\beta_i)}V_{ci}$. When $x\in\mathcal{VI}$, the closed-loop system \eqref{12}-\eqref{36} reduces to $\Dot{x}=-\gamma(x-x_d)$. Thus, we can conclude that the destination is locally exponentially stable and almost globally asymptotically stable using the property \eqref{prop}.\\
% \end{comment}
Finally, we show the quasi-optimality of our obstacle avoidance maneuver. Since the control input \eqref{36} is a composition of the projection from Lemma \ref{lem1}, which generates the shortest path for a considered obstacle, according to Lemma \ref{lem6}, the trajectory $x(t)$ of the closed-loop system \eqref{12}-\eqref{36} generates a quasi-optimal obstacle avoidance maneuver, for $\varepsilon=0$, as per Definition \ref{def2}.
%%%%%%%%%%%%%%%%%%%%%%%%%%%%%%%%%%%%%%%%%%%%%%%%%%%%%%%%%%%%%%%%%

%\section*{ACKNOWLEDGMENT}

%%%%%%%%%%%%%%%%%%%%%%%%%%%%%%%%%%%%%%%%%%%%%%%%%%%%%%%%%%%%%%%%%%%%%%%%%%%%%%%%%

\addtolength{\textheight}{-12cm}   % This command serves to balance the column lengths

%%%%%%%%%%%%%%%%%%%%%%%%%%%%%%%%%%%%%%%%%%%%%%%%%%%%%%%%%%%%%%%%%%%%%%%%%%%%%%%%
%%%%%%%%%%%%%%%%%%%%%%%%%%%%%%%%%%%%%%%%%%%%%%%%%%%%%%%%%%%%%%%%%%%%%%%%%%%%%%%%
\bibliographystyle{ieeetr}
\bibliography{references}

\end{document}